\documentstyle[aps,pre,eqsecnum,multicol,epsf,rotate]{revtex}

\begin{document}
\draft

\title{Critical behavior of $\bbox{O(n)}$--symmetric systems with
	reversible mode--coupling terms: \\
	Stability against detailed--balance violation}

\author{Uwe C. T\"auber}

\address{Department of Physics --- Theoretical Physics, 
	University of Oxford, 1 Keble Road, Oxford OX1 3NP, U.K. \\
	and Linacre College, Oxford OX1 3JA, U.K.}

\author{Zolt\'an R\'acz}

\address{Institute for Theoretical Physics, E\"otv\"os University,
	1088 Budapest, Puskin u. 5--7, Hungary}

\date{\today}
\maketitle

\begin{abstract}
We investigate nonequilibrium critical properties of $O(n)$--symmetric
models with reversible mode--coupling terms. 
Specifically, a variant of the model of Sasv\'ari, Schwabl, and
Sz\'epfalusy is studied, where violation of detailed balance is 
incorporated by allowing the order parameter and the dynamically
coupled conserved quantities to be governed by heat baths of different
temperatures $T_S$ and $T_M$, respectively. 
Dynamic perturbation theory and the field--theoretic renormalization
group are applied to one--loop order, and yield two new fixed points
in addition to the equilibrium ones.
The first one corresponds to $\Theta = T_S / T_M = \infty$ and leads
to model A critical behavior for the order parameter and to anomalous
noise correlations for the generalized angular momenta; the second one
is at $\Theta = 0$ and is characterized by mean--field behavior of the
conserved quantities, by a dynamic exponent $z = d / 2$ equal to that
of the equilibrium SSS model, and by modified static critical
exponents.
However, both these new fixed points are unstable, and upon
approaching the critical point detailed balance is restored, and the
equilibrium static and dynamic critical properties are recovered.
\end{abstract}


\begin{multicols}{2}

\section{Introduction}
 \label{introd}

Nonequilibrium steady states (NESS) have been much investigated, the
main goal being the discovery of their common and distinguishing
features as compared to equilibrium states. A promising approach to
this problem is the study of phase transitions: Since equilibrium 
critical phenomena display a large degree of universality, it is
natural to ask to what extent these universal features remain 
characteristic of nonequilibrium phase transitions.

The basic complication with NESS is that, in addition to the
interactions which entirely define the equilibrium properties, the
dynamics is also essential in determining the steady state
properties. Thus, for example, a classification of nonequilibrium
phase transitions requires not only the understanding of the role of
symmetries of the order parameter, the range of interactions and the
dimensionality of the system, but the clarification of both the
relevance of conservation laws imposed by dynamical symmetries and the
range of the dynamical processes. Possibly, new dimensionality and
anisotropy effects in the dynamics may also be important.

The most frequently studied models with nonequilibrium phase
transitions are generalizations of systems with model A type dynamics
\cite{hohhal}. The transitions in these systems have been shown to be
robust against local nonequilibrium perturbations which do not
conserve the order parameter \cite{grins1} and, remarkably, this
robustness was found to persist even if the {\em dynamical}
perturbations broke the discrete symmetry of the system \cite{kevbea}.
Both locality and the nonconserving character of the perturbations are
essential for the phase transition to stay in the Ising universality
class. Indeed, nonlocal nonequilibrium dynamics generates effective
long--range forces and thus changes the universality class
dramatically \cite{drtbra}. 

The nonequilibrium generalizations of model B type dynamics (with
conserved order parameter) are more interesting. External fields
or local, {\em anisotropic}, nonequilibrium perturbations may drive
the system into a NESS with phase transitions which are not
characterized by any known equilibrium universality class
\cite{bearev}, or belong to universality classes with long--range
interactions \cite{schmit,kevzol}.

Nonequilibrium generalizations of the case when a nonconserved order 
parameter is coupled to a conserved quantity have been considered in
Ref.~\cite{grins2} where it was found that linear coupling to a
conserved quantity generates power--law correlations for the order
parameter. This suggests that, in this situation, long--range
effective interactions are generated in the system, which in turn
govern the critical behavior at the phase transition.

There are several other nonequilibrium phase transitions which have
been studied without considering any equilibrium context. Most notable
among these are phase transitions associated with the presence of an
absorbing state (directed percolation) \cite{absorb}, and the
roughening transition in surface growth and equivalent models such as
the Kardar--Parisi--Zhang equation \cite{kapezt}.

In this paper we continue the investigation of nonequilibrium
generalizations of models originally proposed to describe equilibrium
critical dynamics. Our aim is to study an example where there is a
reversible mode--coupling between the order parameter and another
(conserved) field, using the field--theoretic dynamic renormalization
group (RG) \cite{janded,bajawa}. A simple example of this type of
systems is the Heisenberg model for isotropic ferromagnets where
precession terms introduce a coupling among the different spin
components (model J according to the classification in
Ref.~\cite{hohhal}; for early RG studies of this model see
Ref.~\cite{modelj}; a comprehensive review of the critical dynamics of
ferromagnets is given in Ref.~\cite{erwrev}). However, similar to the
purely relaxational dynamics of models A and B, the effect of a
(spatially isotropic) violation of detailed balance can be removed by
a simple rescaling (see Sec.\ref{modelj}). We shall thus mainly
consider a more complicated model which was originally introduced by
Sasv\'ari, Schwabl, and Sz\'epfalusy in the context of structural
phase transitions \cite{sssmod}. This SSS model consists of a
non--conserved $n$--component order parameter purely dynamically
coupled to the $n (n-1) / 2$ conserved generalized angular momenta
related to the underlying $O(n)$ symmetry of the system. The $n = 2$
realization describes the critical dynamics of planar ferromagnets and
superfluid Helium 4 \cite{modele} (for reviews regarding dynamic
critical phenomena in superfluid Helium, see Ref.~\cite{helium}),
while the case $n = 3$ corresponds to the dynamics of isotropic
antiferromagnets \cite{modelg}. The SSS model, with its dynamic
exponent $z = d / 2$ (below the upper critical dimension $d_c = 4$)
thus encompasses models E and G (according to Ref.~\cite{hohhal}) as
special cases.

We shall generalize the previous field--theoretic RG studies of the
SSS model \cite{sssfth,oerjan} to a nonequilibrium situation by
assuming that the order parameter components $S^\alpha$ and conserved
angular momenta $M^{\alpha \beta}$ are attached to heat baths of 
{\em different} temperatures $T_S$ and $T_M$, respectively. Thus the
detailed--balance condition required for near--equilibrium dynamics is
violated and the flow of energy between the two heat baths ensures
that the steady state is out of equilibrium. The introduction of two
temperatures leads to an additional variable in the problem, namely
the temperature ratio $\Theta = T_S / T_M$ (of which no analog can be
constructed for model J). By studying the RG flow equations to
one--loop order (first order in $\epsilon = 4 - d$), we find 
{\em two new fixed points} corresponding to the cases $\Theta = 0$ and
$\Theta = \infty$, respectively, in addition to the equilibrium fixed
points of the SSS model. The latter are (i) the usual Gaussian fixed
point (describing static and dynamic mean--field behavior, $z = 2$), 
(ii) the model A fixed point, corresponding to a decoupling of the
conserved fields from the order parameter, with the nontrivial static
exponents of the $O(n)$--symmetric $\bbox{\phi}^4$ model and dynamic
exponent $z = 2 + {\cal O}(\epsilon^2)$, and the three nontrivial SSS
dynamic fixed points consisting of the two so--called 
{\em weak--scaling} fixed points with the order parameter and
conserved quantities fluctuating on different time scales,
characterized by the exponents (iii) 
$z_S = 2 - 2 (n-1) \epsilon / (2 n - 1) + {\cal O}(\epsilon^2)$, 
$z_M = 2 - \epsilon / (2n-1) + {\cal O}(\epsilon^2)$, and (iv)
$z_S = 2$ and $z_M = d - 2$, and finally (v) the {\em strong--scaling}
fixed point with $z_S = z_M = z = d / 2$. The results for (iv) and (v)
actually hold to all orders in $\epsilon$, and follow from the exact
sum rule $z_S + z_M = d$ \cite{modele,sssfth,uwedis} (see
Sec.~\ref{genres}). Stability analysis shows that to one--loop order
only the strong--scaling fixed point (v) is stable; however, at least
for $n = 2$ the actual fixed--point values are rather close to its
stability boundary, which allows for the possibility that in fact for
superfluid Helium strong scaling may be violated at the Lambda
transition \cite{sssfth} (a two--loop study of model F, combined with
Borel--resummation techniques, actually suggests the stability of a
weak--scaling fixed point \cite{modelf}).

The stability of the above fixed points may change in a nonequilibrium
situation where the order parameter and conserved variables are
allowed to fluctuate at different temperatures, which explicitly
introduces different characteristic time scales. Indeed, while one of
the two new nonequilibrium fixed points, corresponding to (a) 
$\Theta = \infty$, is described by model A dynamics $z_S = z_M = 2$
(with the usual $\bbox{\phi}^4$ model statics), albeit accompanied by
{\em anomalous noise correlations for the conserved fields}, the
second new fixed point, characerized by (b) $\Theta = 0$, yields,
actually to all orders in $\epsilon$, $z_S = d / 2$ for the order
parameter as in equilibrium, but $z_M = 2$, i.e.: ordinary diffusion
for the generalized angular momenta  (note that the above--mentioned
equilibrium sum rule does not hold here); this unusual behavior is
supplemented by {\em anomalous order parameter noise correlations},
and even {\em modified static critical exponents}. However, stability
analysis reveals that in fact both these fixed points (a) and (b) are
unstable, and for any initial value of $0 < \Theta < \infty$ the flow
asymptotically leads to the {\em stable strong--scaling equilibrium
fixed point} of the SSS model (see Sec.~\ref{looprg}). Thus we
conclude that while violation of detailed balance might be conceived
as a relevant perturbation, in fact the underlying $O(n)$ symmetry in
conjunction with spatial isotropy and the growing correlation length
as the phase transition is approached, effectively {\em restores
detailed balance} (described by the fixed point with $\Theta = 1$),
and thus asymptotically yield the usual static and dynamic critical
behavior of the equilibrium SSS model.

This paper is organized as follows. In the following
Sec.~\ref{dynmod}, we briefly review the derivation of Langevin
equations describing the critial dynamics of $O(n)$--symmetric models
including reversible mode--coupling terms, and consider the possible
relevance of detailed--balance violation for the relaxational models A
and B, as well as model J and the SSS model. Sec.~\ref{neqsss} will
then be devoted to the RG study of the nonequilibrium SSS model as
outlined above, starting with stating some general exact relations and
Ward identities, followed by a detailed study of the one--loop
perturbation theory, the ensuing flow equations, and a discussion of
the physical content and stability of the RG fixed points. Finally, in
Sec.~\ref{sumcon} we shall summarize our results again, draw some
conclusions and provide an outlook on possible future research along
the path followed in this paper. In the Appendix, we provide a list of
the explicit results to one--loop order for the two--, three--, and
four--point functions required for the renormalization of the
nonequilibrium SSS model.

\section{Critical dynamics of models with reversible mode--coupling
	terms} 
 \label{dynmod}

\subsection{General considerations}
 \label{gencon}

The universal static critical behavior of a system which is invariant
with respect to rotations of its $n$--component order parameter and
displays a second--order phase transition is described by the
following $O(n)$--symmetric $\bbox{\phi}^4$ Landau--Ginzburg--Wilson
hamiltonian in $d$ space dimensions
\begin{eqnarray}
	H[\{ S_0^\alpha \}] = \int \! d^dx \Biggl\{
	&&{r_0 \over 2} \sum_{\alpha=1}^n S_0^\alpha({\bf x})^2 +	
	{1 \over 2} \sum_{\alpha=1}^n \left[ 
	\bbox{\nabla} S_0^\alpha({\bf x} \right)]^2 \nonumber \\
	&&\quad + {u_0 \over 4 !} \left[ 
	\sum_{\alpha=1}^n S_0^\alpha({\bf x})^2 \right]^2 \Biggr\} \ ;
 \label{hamilt}
\end{eqnarray}
here $r_0 = (T - T_c^0) / T_c^0$ denotes the relative distance from
the mean--field critical temperature $T_c^0$, and we denote
unrenormalized quantities by a subscript ``$0$''. This effective free
energy determines the equilibrium probability distribution for the
vector order parameter $S_0^\alpha$,
\begin{equation}
	P_{\rm eq}[\{ S_0^\alpha \}] = 
	{e^{-H[\{ S_0^\alpha \}] / k_{\rm B} T} \over 
	\int {\cal D}[\{ S_0^\alpha \}] 
		e^{-H[\{ S_0^\alpha \}] / k_{\rm B} T}} \ ,
 \label{eqdist}
\end{equation}
and furthermore provides the starting point for the construction of
the field--theoretic static renormalization group which by virtue of a
perturbation (loop) expansion in the nonlinearity $u_0$ provides a
systematic means to compute the two independent static critical
exponents $\eta$ and $\nu$ either in an $\epsilon$ expansion about the
upper critical dimension $d_c = 4$, or directly in fixed
dimensionality $d$ \cite{stafth}. Here, $\eta$ is the anomalous
dimension which describes the power--law decay of the order parameter
correlations at the critical point, 
$\langle S^\alpha({\bf x}) S^\beta({\bf x}') \rangle \propto  
1 / |{\bf x} - {\bf x}'|^{d - 2 + \eta}$, and the exponent $\nu$
characterizes the divergence of the correlation length as $T_c$ is
approached, $\xi \propto |T - T_c|^{- \nu}$.

The simplest dynamics that may be imposed on the order parameter
fluctuations $S_0^\alpha({\bf x},t)$ in order to describe how the
system relaxes to equilibrium (for which the mean--field stationarity
condition $\delta H[\{ S_0^\alpha \}] / \delta S_0^\alpha = 0$ holds)
is then given by the following Langevin--type equations of motion
\begin{equation}
	{\partial S_0^\alpha({\bf x},t) \over \partial t} = 
	- \lambda_0 (i \bbox{\nabla})^a \, 
	{\delta H[\{ S_0^\alpha \}] \over 
	\delta S_0^\alpha({\bf x},t)} + \zeta^\alpha({\bf x},t) \ ,
 \label{relmod}
\end{equation}
where the temporal average of the stochastic forces is assumed to
vanish, $\langle \zeta^\alpha({\bf x},t) \rangle = 0$. In equilibrium,
furthermore an Einstein relation connects the second moment of the
uncorrelated (white) noise with the relaxation coefficient,
\begin{equation}
	\langle \zeta^\alpha({\bf x},t) \zeta^\beta({\bf x}',t') 
	\rangle = 2 \lambda_0 k_{\rm B} T (i \bbox{\nabla})^a
		\delta({\bf x} - {\bf x}') \delta( t - t') 
				\delta^{\alpha \beta} ;
 \label{abnois}
\end{equation}
this ensures that the probability distribution $P[\{ S_0^\alpha \}]$
finally approaches the equilibrium distribution (\ref{eqdist}) in the
limit $t \to \infty$, as can be readily checked with the aid of the
associated Fokker--Planck equation. Eq.~(\ref{relmod}) incorporates
both the case of a nonconserved order parameter with purely
relaxational dynamics ($a = 0$) and a conserved order parameter which
as a consequence of the ensuing continuity equation relaxes
diffusively ($a = 2$). In the classification scheme of Hohenberg and
Halperin, these situations are referred to as models A and B,
respectively, and the corresponding dynamic critical exponents
describing the critical slowing down near the phase transition
(characteristic timescales diverge as 
$t_c \propto \xi^z \propto |T - T_c|^{- z \nu}$) are given
in terms of the static exponent $\eta$ by $z = 4 - \eta$ (model B)
and $z = 2 + c \eta$ (model A) \cite{hohhal}. In the latter case,
however, $c$ is a new universal number and therefore $z$ is an
independent exponent not determined by the static critical exponents.

One may already anticipate that an {\em isotropic} violation of the
Einstein relation (\ref{abnois}), which is a consequence of an
underlying detailed balance condition, by choosing a coefficient
${\tilde \lambda}_0$ instead of $\lambda_0 k_{\rm B} T$ for the noise
correlator, merely amounts to a change in the order parameter
temperature $T$. Therefore, as long as one remains sufficiently close
to the critical point, the {\em universal} critical behavior
(exponents, amplitude ratios, etc.) will not be affected, while only
nonuniversal amplitudes become modified through a rescaled nonlinear
coupling $u_0$ (see Sec.~\ref{modelj} and Ref.~\cite{grins1}).

However, in an $O(n)$--symmetric system there are always additional
slow diffusive modes present. In our case these modes are associated
with the conserved generalized angular momenta $M_0^{\alpha\beta}$
which generate the rotations in order parameter space. Generically,
they couple to the order parameter fluctuations, and therefore
Eq.~(\ref{relmod}) does not correctly describe their dynamics. Two
cases can now be distinguished: (i) The vector order parameter itself
is identical with the generators of the group $O(n)$; this yields, for
$n = 3$, precisely the dynamics of isotropic Heisenberg ferromagnets
\cite{modelj}, model J according to Ref.~\cite{hohhal}. (ii) The order
parameter is nonconserved, and the conserved angular momenta
constitute new dynamical variables; this defines the $O(n)$--symmetric
model introduced by Sasv\'ari, Schwabl, and Sz\'epfalusy
\cite{sssmod}, and encompasses both model E for the dynamics of the XY
model, i.e., of planar ferromagnets and superfluid Helium 4 ($n = 2$)
\cite{modele}, and model G for isotropic antiferromagnets ($n = 3$)
\cite{modelg}.

Upon collecting the order parameter and angular momentum components in
a large vector $\psi^\alpha = (S^\alpha , M^{\alpha \beta})$, the
general structure of the ensuing Langevin equations reads 
\cite{modelj,sssmod}
\begin{equation}
	{\partial \psi^\alpha({\bf x},t) \over \partial t} =
	V^\alpha[\{ \psi^\alpha \}]({\bf x},t) - L^\alpha 
	{\delta H[\{ \psi^\alpha \}] \over 
	\delta \psi^\alpha({\bf x},t)} + \zeta^\alpha({\bf x},t) \ , 
 \label{genlan}
\end{equation}
where $L^\alpha = \lambda$ or $L^\alpha = - D \bbox{\nabla}^2$ for all
the nonconserved and conserved fields, respectively. The second term
on the right--hand side of Eq.~(\ref{genlan}) describes irreversible
relaxation processes as in models A and B [Eq.~(\ref{relmod})]; the
first term, on the other hand, consists of {\em reversible}
``mode--couplings'', which are given entirely by the Poisson brackets 
$Q^{\alpha \beta}[ \{ \psi^\alpha \}] \propto 
\{ \psi^\alpha , \psi^\beta \}$. As can be shown with the
Kawasaki--Mori--Zwanzig projector formalism, $V[\{ \psi^\alpha \}]$
assumes the form of a ``streaming velocity'' in the space of the
$\psi^\alpha$, namely 
\begin{equation}
	V^\alpha[\{ \psi^\alpha \}] = g \sum_\beta \left( 
	k_{\rm B} T {\delta Q^{\alpha\beta} \over \delta \psi^\beta} -
	Q^{\alpha\beta} {\delta H[\{ \psi^\alpha \}] \over 
					\delta \psi^\beta} \right) \ .
 \label{genmct}
\end{equation}
Note that the mode--coupling constants $g$ are {\em independent} of
$\alpha$, which guarantees that $V^\alpha[\{ \psi^\alpha \}] 
e^{- H[\{ \psi^\alpha \}] / k_{\rm B} T}$ is divergence--free,
\begin{equation}
	\sum_\alpha {\delta \over \delta \psi^\alpha} 
	\left( V^\alpha[\{ \psi^\alpha \}] 
	e^{- H[\{ \psi^\alpha \}] / k_{\rm B} T} \right) = 0 \ , 
 \label{divcon}
\end{equation}
and therefore the equilibrium distribution 
$P_{\rm eq}[\{ \psi^\alpha \}] \propto 
e^{- H[\{ \psi^\alpha \}] / k_{\rm B} T}$ is not affected by the
mode--coupling terms which are of purely dynamical origin.

We defer the explicit construction of the mode--coupling terms for
model J and the SSS model to the following subsections, and close this
general discussion with a brief outline how one may construct an
effective field theory from Langevin equations of the type
\begin{equation}
	{\partial \psi^\alpha({\bf x},t) \over \partial t} = 
		K^\alpha[\{ \psi^\alpha \}]({\bf x},t) + 
				\zeta^\alpha({\bf x},t) \ ,
 \label{laneqs}
\end{equation}
see Eq.~(\ref{genlan}), with 
$\langle \zeta^\alpha({\bf x},t) \rangle = 0$ and the general noise
correlator
\begin{equation}
	\langle \zeta^\alpha({\bf x},t) \zeta^\beta({\bf x}',t') 
	\rangle = 2 L^\alpha \delta({\bf x} - {\bf x}') \delta(t - t')
					\delta^{\alpha \beta} \ ,
 \label{lannoi}
\end{equation}
see Refs.~\cite{janded,bajawa}. This form of the white noise may be
inferred from a Gaussian distribution for the stochastic forces
\begin{equation}
	W[\{ \zeta^\alpha \}] \propto \exp 
	\left[ - {1 \over 4} \int \! d^dx \int \! dt \sum_\alpha 
		\zeta^\alpha (L^\alpha)^{-1} \zeta^\alpha \right] \ ;
 \label{noidis}
\end{equation}
eliminating $\zeta^\alpha$ via Eq.~(\ref{laneqs}) then immediately
yields the desired probability distribution for the fields
$\psi^\alpha$,
\begin{equation}
	W[\{ \zeta^\alpha \}] {\cal D}[\{ \zeta^\alpha \}] =
	P[\{ \psi^\alpha \}] {\cal D}[\{ \psi^\alpha \}] \propto
	e^{G[\{ \psi^\alpha \}]} {\cal D}[\{ \psi^\alpha \}] ,
 \label{prodis}
\end{equation}
with the Onsager--Machlup functional
\begin{eqnarray}
	G[\{ \psi^\alpha \}] = - {1 \over 4} &&\int \! d^dx \int \! dt 
	\sum_\alpha \left( {\partial \psi^\alpha \over \partial t} - 
	K^\alpha[\{ \psi^\alpha \}] \right) \times \nonumber \\
	&&\times (L^\alpha)^{-1} \left( {\partial \psi^\alpha \over 
		\partial t} - K^\alpha[\{ \psi^\alpha \}] \right) \ . 
 \label{onsmlp}
\end{eqnarray}

From this functional one could already construct a perturbation
expansion for correlation functions of the fields $\psi^\alpha$;
however, as for conserved quantities the inverse of the Onsager
coefficient $L^\alpha$ is singular, and furthermore high
nonlinearities $\propto K^\alpha[\{ \psi^\alpha \}]^2$ appear, it is
convenient to introduce Martin--Siggia--Rose auxiliary fields via a
Gaussian transformation to partially linearize the above functional. 
This leads to
\begin{equation}
	P[\{ \psi^\alpha \}] \propto 
	\int {\cal D}[\{ i {\tilde \psi}^\alpha \}] 
	e^{J[\{ {\tilde \psi}^\alpha \} , \{ \psi^\alpha \}]} \ ,
 \label{genfun}
\end{equation}
with the Janssen--De~Dominicis functional
\begin{eqnarray}
	J[\{ {\tilde \psi}^\alpha \} , \{ \psi^\alpha \}] = 
	&&\int \! d^dx \int \! dt \sum_\alpha \Biggl[ 
	{\tilde \psi}^\alpha L^\alpha {\tilde \psi}^\alpha - \nonumber
	\\ &&- {\tilde \psi}^\alpha \left( {\partial \psi^\alpha \over 
	\partial t} - K^\alpha[\{ \psi^\alpha \}] \right) \Biggr] \ .
 \label{janded}
\end{eqnarray}

Eq.~(\ref{janded}) will provide the starting point for our discussion
of the nonequilibrium dynamics of the isotropic ferromagnet (model J)
as well as that of the SSS model in the subsequent subsections. In
Sec.~\ref{neqsss}, we shall use the corresponding
Janssen--De~Dominicis functional for the construction of the dynamical
field theory of the SSS model with broken detailed balance, and
therefrom infer its RG flow equations. We finally remark that both in
Eqs.~(\ref{onsmlp}) and (\ref{janded}) we have omitted contributions
stemming from the functional determinant
${\cal D}[\{ \zeta^\alpha \}] / {\cal D}[\{ \psi^\alpha \}]$. As is
shown in Refs.~\cite{bajawa,sssfth}, these terms precisely cancel any
{\em acausal} Feynman diagrams for the dynamic response function that
could be constructed from the above functionals; and upon restricting
the perturbation expansion to those contributions which are consistent
with causality requirements, we may therefore safely neglect these
additional terms.

\subsection{Model J -- isotropic ferromagnets}
 \label{modelj}

We now turn explicitly to the construction of the Langevin equation
for the critical dynamics of isotropic ferromagnets \cite{modelj}. In
this case, $n = 3$, and the order parameter consists of the three spin
components $S^x$, $S^y$, and $S^z$. The total magnetization is a
conserved quantity (hence $a = 2$), and in fact the $S^\alpha$ are
identical with the generators of the rotation group $O(n)$: 
$M^{12} = S^z$, $M^{23} = S^x$, and $M^{13} = - S^y$. The Poisson
brackets between the spin components read
\begin{equation}
	\left\{ S^\alpha , S^\beta \right\} = 
	\sum_\gamma \epsilon^{\alpha \beta \gamma} S^\gamma \ ,
 \label{ferpbr}
\end{equation}
which immediately yields the streaming velocity
\begin{eqnarray}
	V^\alpha[\{ S^\alpha \}] &=& - g \sum_{\beta,\gamma} 
		\epsilon^{\alpha \beta \gamma} S^\gamma 
	{\delta H[\{ S^\alpha \}] \over \delta S^\beta} \nonumber \\
	&=& - g \sum_{\beta,\gamma} \epsilon^{\alpha \beta \gamma} 
			S^\beta \bbox{\nabla}^2 S^\gamma \ ,
 \label{fermct}
\end{eqnarray}
for the contractions of the fully antisymmetric tensor 
$\epsilon^{\alpha \beta \gamma}$ with all the symmetric terms in 
Eq.~(\ref{hamilt}) vanish, leaving only the contribution stemming from
the gradient term in the hamiltonian. The mode--coupling terms
(\ref{fermct}) represent the spin precession in the effective field
generated by the other spins, and in the ordered phase leads to
propagating spin waves (Goldstone modes) with quadratic dispersion
$\omega(q) \propto q^2$.

The complete Langevin equation for the conserved order parameter of
isotropic ferromagnets (model J according to Ref.~\cite{hohhal})
finally reads 
\begin{equation}
	{\partial S_0^\alpha \over \partial t} = 
	- g_0 \sum_{\beta,\gamma} \epsilon^{\alpha \beta \gamma} 
		S_0^\beta \bbox{\nabla}^2 S_0^\gamma +
	\lambda_0 \bbox{\nabla}^2 {\delta H[\{ S_0^\alpha\}] \over
		\delta S_0^\alpha} + \zeta^\alpha \ , 
 \label{ferlan}
\end{equation}
with $\langle \zeta^\alpha({\bf x},t) \rangle = 0$ and
\begin{equation}
	\langle \zeta^\alpha({\bf x},t) \zeta^\beta({\bf x}',t')
	\rangle = - 2 {\tilde \lambda}_0 \bbox{\nabla}^2 
		\delta({\bf x} - {\bf x}') \delta( t - t') 
					\delta^{\alpha \beta} \ .
 \label{fernoi}
\end{equation}
Here we have already allowed for a violation of the detailed--balance
condition by introducing a noise strength ${\tilde \lambda}_0$ that is
not necessarily equal to $\lambda_0 k_{\rm B} T$, where $\lambda_0$ is
the spin diffusion constant. However, the form of Eqs.~(\ref{ferlan})
and (\ref{fernoi}) already suggests that similar to the case of the
purely relaxational models A and B, the ratio ${\tilde \lambda}_0 /
\lambda_0$ may be absorbed into a rescaled temperature $T$, and
modified nonlinear couplings $u_0$ and $g_0$. 

This can be readily seen by employing the corresponding
Janssen--De~Dominicis functional (\ref{janded}); for our
nonequilibrium model J this becomes a sum of the dynamic functionals
for the relaxational models
\begin{eqnarray}
	&&J_{\rm rel}[\{ {\tilde S}_0^\alpha \} , \{ S_0^\alpha \}] = 
	\int \! d^dx \int \! dt \sum_\alpha \Biggl\{ 
	{\tilde \lambda}_0 {\tilde S}_0^\alpha (i \bbox{\nabla})^a 
		{\tilde S}_0^\alpha - \nonumber \\ &&\qquad \qquad - 
	{\tilde S}_0^\alpha \left[ {\partial \over \partial t} + 
	\lambda_0 (i \bbox{\nabla})^a \left( r_0 - 
	\bbox{\nabla}^2 \right) \right] S_0^\alpha - \nonumber \\ 
	&&\qquad \qquad - \lambda_0 {u_0 \over 6} 
		\sum_\beta {\tilde S}_0^\alpha (i \bbox{\nabla})^a 
			S_0^\alpha S_0^\beta S_0^\beta \Biggr\} \ , 
 \label{abfunc}
\end{eqnarray}
with $a = 2$, and the additional contribution stemming from the
reversible spin precession term,
\begin{equation}
	J_{\rm mc}[\{ {\tilde S}_0^\alpha \},\{ S_0^\alpha \}] = 
	- g_0 \int \! d^dx \int \! dt \! \sum_{\alpha,\beta,\gamma} 
		\epsilon^{\alpha \beta \gamma}
	{\tilde S}_0^\alpha S_0^\beta \bbox{\nabla}^2 S_0^\gamma \ .
 \label{fmfunc}
\end{equation}
Rescaling the fields according to
\begin{equation}
	{\tilde S}_0^\alpha \to \left( {\lambda_0 \over 
	{\tilde \lambda}_0} \right)^{1/2} {\tilde S}_0^\alpha \; , 
	\quad S_0^\alpha \to \left( {{\tilde \lambda}_0 \over 
				\lambda_0} \right)^{1/2} S_0^\alpha
 \label{opscal}
\end{equation}
then renders the noise strength and the relaxation constant in the
quadratic part (first and second line) of Eq.~(\ref{abfunc}) equal,
and if in addition the rescaled static and dynamic nonlinear couplings
\begin{equation}
	{\tilde u}_0 = {{\tilde \lambda}_0 \over \lambda_0} \, u_0 \;
	, \quad {\tilde g}_0 = \left( {{\tilde \lambda}_0 \over 
				\lambda_0} \right)^{1/2} g_0
 \label{abscal}
\end{equation}
are introduced, the ensuing Janssen--De~Dominicis functionals for the
above nonequilibrium generalizations of the relaxational models as
well as model J appear in precisely the same form as in equilibrium
where detailed balance holds. As both the renormalized counterparts of
${\tilde u}_0$ and ${\tilde g}_0 / \lambda_0^2$ approach universal
fixed--point values near the transition, the modifications in
Eq.~(\ref{abscal}) merely enter {\em nonuniversal} amplitudes. It is
therefore established that the {\em critical} properties of neither
the relaxational models A and B nor isotropic ferromagnets (model J)
are affected by violating the detailed--balance condition. It is,
however, important to note that both the $O(n)$ symmetry {\em and} the
spatial isotropy of the models have been left intact by the above
nonequilibrium generalization. For the dynamics of Heisenberg
ferromagnets, we finally remark that the dynamic critical exponent
becomes $z = (d + 2 - \eta) / 2$, as a consequence of a Ward identity
stemming from the underlying $O(3)$ symmetry (see also
Sec.~\ref{genres}). Further details regarding the dynamic critical
behavior of ferromagnets may be found in Ref.~\cite{erwrev}.

\subsection{The SSS model -- planar ferromagnets, isotropic
		antiferromagnets} 
 \label{modsss}

More interesting for the issue of violating detailed balance will
clearly be a situation where there are {\it two} independent
temperature scales conceivable, and therefore a simple temperature
rescaling will not suffice to render the field theory identical to the
equilibrium one. We therefore consider a nonequilibrium version of the
$O(n)$--symmetric SSS model, where a nonconserved $n$--component order
parameter couples to $n (n-1) / 2$ conserved generalized angular
momenta \cite{sssmod}; possible realizations of this are (i) for 
$n = 2$: the critical dynamics of the XY model \cite{modele} (model E
according to Ref.~\cite{hohhal}), with the order parameter components
$S^x$ and $S^y$, and the conserved quantity $M^{12} = S^z$, which
generates rotations in the $xy$--plane; (ii) for $n = 3$: the dynamic
critical behavior of isotropic antiferromagnets, with $S^x$, $S^y$,
and $S^z$ representing the components of the staggered magnetization,
and $M^{12} = M^z$, $M^{23} = M^x$, and $M^{13} = - M^y$ denoting the
components of the magnetization itself, which are conserved and
can be identified with the generators of $O(3)$ (model G
\cite{modelg}). 

The variables $M_0^{\alpha \beta}$ are noncritical quantities, and
their coupling to the order parameter fluctuations $S_0^\alpha$ is of
purely dynamical character. Hence it suffices to simply add a
quadratic term to the hamiltonian (\ref{hamilt}),
\begin{equation}
	H[\{ S_0^\alpha \} , \{ M_0^{\alpha \beta} \}] = 
	H[\{ S_0^\alpha \}] + \int \! d^dx {1 \over 2} 
	\sum_{\alpha > \beta} M_0^{\alpha \beta}({\bf x})^2 \ ;
 \label{stasss}
\end{equation}
and for the construction of the reversible mode--coupling terms, again
all that is required are the following Poisson brackets,
\begin{eqnarray}
	\left\{ S^\alpha , S^\beta \right\} = 0 \; &&, \quad
	\left\{ M^{\alpha \beta} , S^\gamma \right\} = 
	\delta^{\alpha \gamma} S^\beta -
		\delta^{\beta \gamma} S^\alpha \; , \nonumber \\
	\left\{ M^{\alpha \beta} , M^{\gamma \delta} \right\} &&= 
	\delta^{\alpha \gamma} M^{\beta \delta} +
		\delta^{\beta \delta} M^{\alpha \gamma} - \nonumber \\
	&&\quad - \delta^{\alpha \delta} M^{\beta \gamma} -
		\delta^{\beta \gamma} M^{\alpha \delta} \ .
 \label{ssspbr}
\end{eqnarray}
Upon inserting (\ref{ssspbr}) into Eq.~(\ref{genmct}), one readily
finds the following mode--coupling terms in the equations of motion of
the order parameter,
\begin{equation}
	V^\alpha[\{ S^\alpha \} , \{ M^{\alpha \beta} \}] = 
	g \sum_\beta S^\beta {\delta H \over \delta M^{\alpha \beta}}
	= g \sum_\beta M^{\alpha \beta} S^\beta \ ,
 \label{opmcte}
\end{equation}
and in the equation of motion for the conserved angular momenta,
\begin{eqnarray}
	V^{\alpha \beta}[\{ S^\alpha \} , \{ M^{\alpha \beta} \}] &&=
	g \left( S^\alpha {\delta H \over \delta S^\beta} -
	S^\beta {\delta H \over \delta S^\alpha} \right) + \nonumber \\
	+ g \sum_\gamma &&\left( 
	M^{\alpha \gamma} {\delta H \over \delta M^{\beta \gamma}} -
	M^{\beta \gamma} {\delta H \over \delta M^{\alpha \gamma}} 
						\right) \nonumber \\
	= - g \Bigl( S^\alpha \bbox{\nabla}^2 S^\beta &&- 
			S^\beta \bbox{\nabla}^2 S^\alpha \Bigr) \ ,
 \label{ammcte}
\end{eqnarray}
respectively. Note that as for model J [Eq.~(\ref{fermct})], here as a
consequence of the antisymmetry of the Poisson brackets only the
gradient terms in the hamiltonian contribute. We remark that in the
ordered phase the above reversible mode couplings produce propagating
Goldstone modes with linear dispersion $\omega(q) \propto q$.

Thus we arrive at the following set of coupled nonlinear Langevin
equations that define the SSS model,
\begin{eqnarray}
	{\partial S_0^\alpha \over \partial t} &=& 
	g_0 \sum_\beta M_0^{\alpha \beta} S_0^\beta - 
	\lambda_0 {\delta H[\{ S_0^\alpha\}] \over \delta S_0^\alpha}
	+ \zeta^\alpha \ ,
 \label{soplan} \\
	{\partial M_0^{\alpha \beta} \over \partial t} &=& 
	- g_0 \left( S_0^\alpha \bbox{\nabla}^2 S_0^\beta -
	S_0^\beta \bbox{\nabla}^2 S_0^\alpha \right) + \nonumber \\
	&&\qquad + D_0 \bbox{\nabla}^2 M_0^{\alpha \beta} + 
					\eta^{\alpha \beta} \ ,
 \label{samlan}
\end{eqnarray}
with $\langle \zeta^\alpha({\bf x},t) \rangle = 0$, 
$\langle \eta^{\alpha \beta}({\bf x},t) \rangle = 0$, and
\begin{eqnarray}
	\langle \zeta^\alpha({\bf x},t) \zeta^\beta({\bf x}',t') 
	\rangle &=& 2 {\tilde \lambda}_0 \delta({\bf x} - {\bf x}') 
			\delta( t - t') \delta^{\alpha \beta} \ ,
 \label{sopnoi} \\
	\langle \eta^{\alpha \beta}({\bf x},t) 
	\eta^{\gamma \delta}({\bf x}',t') \rangle &=& 
	- 2 {\tilde D}_0 \bbox{\nabla}^2 \delta({\bf x} - {\bf x}') 
				\delta( t - t') \times \nonumber \\
	&&\times \left( \delta^{\alpha \beta} \delta^{\gamma \delta} -
	\delta^{\alpha \delta} \delta^{\beta \gamma} \right) \ .
 \label{samnoi}
\end{eqnarray}
Here we have allowed for violation of detailed balance via introducing
noise coefficients ${\tilde \lambda}_0$ and ${\tilde D}_0$ which are
in general {\em not} taken equal to $\lambda_0 k_{\rm B} T_S$ and 
$D_0 k_{\rm B} T_M$, respectively, where $T_S$ and $T_M$ are the
temperatures of the heat baths of the order parameter and of the
conserved variables. Yet we now also have the {\em additional} freedom
to choose the ratio $\Theta_0 = T_S / T_M$ different from $1$, which
corresponds to a violation of the detailed balance in the dynamical
{\em coupling} of the modes $S_0^\alpha$ and $M_0^{\alpha \beta}$. We
should stress again that neither the underlying $O(n)$ symmetry 
{\em nor} the spatial isotropy are affected by this specific
nonequilibrium perturbation.

This becomes clear upon considering the Janssen--De~Dominicis
functional (\ref{janded}) which corresponds to
Eqs.~(\ref{soplan})--(\ref{samnoi}). Its harmonic part now reads
\begin{eqnarray}
	&&J_{\rm har}[\{ {\tilde S}_0^\alpha \} , \{ S_0^\alpha \} ,
	\{ {\tilde M}_0^{\alpha\beta} \} , \{ M_0^{\alpha\beta} \}] = 
							\nonumber \\
	&&= \int \! d^dx \int \! dt \Biggl\{ \sum_\alpha 
	{\tilde \lambda}_0 {\tilde S}_0^\alpha {\tilde S}_0^\alpha - 
						\nonumber \\
	&&\qquad \qquad - \sum_\alpha {\tilde S}_0^\alpha 
	\left[ {\partial \over \partial t} + \lambda_0 \left( r_0 - 
	\bbox{\nabla}^2 \right) \right] S_0^\alpha - \nonumber \\ 
	&&\qquad \qquad - \sum_{\alpha > \beta} {\tilde D}_0 
		{\tilde M}_0^{\alpha \beta} \bbox{\nabla}^2
			{\tilde M}_0^{\alpha \beta} - \nonumber \\
	&&\qquad \qquad - \sum_{\alpha > \beta} 
	{\tilde M}_0^{\alpha \beta} \left( {\partial \over \partial t}
	- D_0 \bbox{\nabla}^2 \right) M_0^{\alpha \beta} \Biggr\} \ , 
 \label{harfun}
\end{eqnarray}
and can be readily rescaled to the equilibrium form via
Eq.~(\ref{opscal}) combined with
\begin{equation}
	{\tilde M}_0^{\alpha \beta} \to \left( {D_0 \over 
	{\tilde D}_0} \right)^{1/2} {\tilde M}_0^{\alpha \beta} \; , 
	\quad M_0^{\alpha \beta} \to \left( {{\tilde D}_0 \over 
		D_0} \right)^{1/2} M_0^{\alpha \beta} \ .
 \label{amscal}
\end{equation}
Thereby, the relaxation vertex
\begin{equation}
	J_{\rm rel}[\{ {\tilde S}_0^\alpha \} , \{ S_0^\alpha \}] = 
	- \lambda_0 {u_0 \over 6} \int \! d^dx \int \! dt 
	\sum_{\alpha,\beta} {\tilde S}_0^\alpha S_0^\alpha 
					S_0^\beta S_0^\beta
 \label{relfun}
\end{equation}
attains the new effective coupling
\begin{equation}
	{\tilde u}_0 = {{\tilde \lambda}_0 \over \lambda_0} \, u_0 \ .
 \label{stscal}
\end{equation}
For the mode--coupling terms,
\begin{eqnarray}
	&&J_{\rm mc}[\{ {\tilde S}_0^\alpha \} , \{ S_0^\alpha \} ,
	\{ {\tilde M}_0^{\alpha\beta} \} , \{ M_0^{\alpha\beta} \}] = 
							\nonumber \\
	&&= \int \! d^dx \int \! dt \sum_{\alpha , \beta} \Biggl\{ 
	g_0 {\tilde S}_0^\alpha M_0^{\alpha \beta} S_0^\beta -
							\nonumber \\
	&&\qquad \qquad - {g_0 \over 2} \, {\tilde M}_0^{\alpha \beta}
	\left( S_0^\alpha \bbox{\nabla} S_0^\beta - 
	S_0^\beta \bbox{\nabla} S_0^\alpha \right) \Biggr\} \ , 
 \label{mctfun}
\end{eqnarray}
however, which originally have identical couplings $g_0$, the effect
of this rescaling procedure is to generate two {\em different}
dynamical coupling constants in the first and second terms of
Eq.~(\ref{mctfun}), respectively, namely
\begin{equation}
	{\tilde g}_0 = \left( {{\tilde D}_0 \over D_0} \right)^{1/2}
	g_0 \; , \quad {\tilde g}_0' = \Theta_0 \, {\tilde g}_0 \ ,
 \label{dyscal}
\end{equation}
where
\begin{equation}
	\Theta_0 = {{\tilde \lambda}_0 \over \lambda_0} \,
		{D_0 \over {\tilde D}_0} \ .
 \label{tratio}
\end{equation}
Thus, even if both equations (\ref{soplan}) and (\ref{samlan}) obey
detailed balance separately, two different dynamic couplings will be
generated as long as $T_S \not= T_M$, and then the new variable 
$\Theta_0 = T_S / T_M$ describes the deviation from equilibrium. With
the two independent couplings ${\tilde g}_0$ and ${\tilde g}_0'$, the
renormalization group equations will become different as compared to
the equilibrium situation, and new critical behavior may be expected
at least in the extreme cases where the temperature ratio is either
$\Theta_0 = 0$ or $\Theta_0 = \infty$. In the following
Sec.~\ref{neqsss}, we shall proceed with a detailed investigation of 
the one--loop flow equations of the nonequilibrium SSS model, as given
by the field theory (\ref{harfun}), (\ref{relfun}), and
(\ref{mctfun}), as function of the couplings (\ref{stscal}),
(\ref{dyscal}), and (\ref{tratio}).

\section{Renormalization of the nonequilibrium SSS model}
 \label{neqsss}

\subsection{Response functions and Ward identities}
 \label{genres}

By adding source terms to the Janssen--De~Dominicis functional
(\ref{janded}), one arrives at the generating functional
\begin{eqnarray}
	&&Z[\{ {\tilde h}^\alpha \} , \{ h^\alpha \}] \propto
	\int \! {\cal D}[\{ i {\tilde \psi}_0^\alpha \}]
	{\cal D}[\{ \psi_0^\alpha \}] \,
	e^{J[\{ {\tilde \psi}_0^\alpha \} , \{ \psi_0^\alpha \}]}
	\times \nonumber \\ 
	&&\qquad \times \exp \int \! d^dx \int \! dt \sum_\alpha 
	\left( {\tilde h}^\alpha {\tilde \psi}_0^\alpha +
		h^\alpha \psi_0^\alpha \right) \ ,
 \label{genrfn}
\end{eqnarray}	
and the $({\tilde N} N)$--point correlation functions (cumulants) 
$G^c_{0 \, {\tilde \psi}^{\tilde N} \psi^N}$ can be obtained from 
$\ln Z$ via functional derivatives with respect to the sources 
${\tilde h}^\alpha$ and $h^\alpha$, and then taking all 
${\tilde h}^\alpha = h^\alpha = 0$. Following the usual
field--theoretic techniques \cite{stafth,bajawa}, we furthermore
define the generating functional for the one--particle irreducible
vertex functions using 
${\tilde \phi}_0^\alpha = \delta \ln Z / \delta {\tilde h}^\alpha$ and
$\phi_o^\alpha = \delta \ln Z / \delta h^\alpha$ via the Legendre
transform
\begin{eqnarray}
	\Gamma[\{ {\tilde \phi}_0^\alpha \} , \{ \phi_0^\alpha \}] &&=
	- \ln Z [\{ {\tilde h}^\alpha \} , \{ h^\alpha \}] + \nonumber
	\\ &&+ \int \! d^dx \int \! dt \sum_\alpha  
	\left( {\tilde h}^\alpha {\tilde \phi}_0^\alpha +
		h^\alpha \phi_0^\alpha \right) \ ;
 \label{vertfn}
\end{eqnarray}
the $({\tilde N} N)$--point vertex functions 
$\Gamma_{0 \, {\tilde \psi}^{\tilde N} \psi^N}$ then follow via
functional derivatives of (\ref{vertfn}) with respect to ${\tilde
\phi}_0^\alpha$ and $\phi_0^\alpha$.

With $\langle \psi_0^\alpha \rangle = 0$ we can write 
$\langle \psi_0^\alpha({\bf x},t) {\tilde \psi}_0^\beta({\bf x}',t') 
\rangle = G^c_{0 \, {\tilde \psi} \psi}({\bf x} - {\bf x}',t - t') 
\delta^{\alpha \beta}$, etc., and upon introducing the Fourier
transform according to $\psi^\alpha({\bf q},\omega) = \int \!d^dx \int
\! dt \, \psi^\alpha({\bf x},t) e^{-i({\bf q}\cdot{\bf x}-\omega t)}$,
one finds the following connections between the two--point correlation
and vertex functions,
\begin{eqnarray}
	G^c_{0 \, {\tilde \psi} \psi}({\bf q},\omega) &=& 
	\Gamma_{0 \, {\tilde \psi} \psi}(-{\bf q},-\omega)^{-1} \ , \\ 
 \label{verfn1}
	G^c_{0 \, \psi \psi}({\bf q},\omega) &=& - \,
	{\Gamma_{0 \,{\tilde \psi}{\tilde \psi}}({\bf q},\omega) \over
	| \Gamma_{0 \, {\tilde \psi}\psi}({\bf q},\omega) |^2} \ .
 \label{verfn2}
\end{eqnarray}
For the ${\tilde N} N$--point functions with ${\tilde N} N > 2$,
relations similar to (\ref{verfn2}) hold, see Eq.~(\ref{compop})
below.

In order to assign a meaning to the auxiliary fields, we compute the
response functions for the SSS model by first adding external fields
to the hamiltonian (\ref{stasss}) \cite{bajawa,uwedis},
\begin{equation}
	H \to H - \int \! d^dx \left[ \sum_\alpha {\tilde h}^\alpha 
	S_0^\alpha + \sum_{\alpha > \beta} {\tilde H}^{\alpha \beta}
	M_0^{\alpha \beta} \right] \ ,
 \label{extfld}
\end{equation}
which produces the following additional terms in the dynamic
functional,
\begin{eqnarray}
	J \to J &&+ \! \int \! d^dx \! \int \! dt \Biggl[ 
	\lambda_0 \sum_\alpha {\tilde h}^\alpha {\tilde S}_0^\alpha -
	D_0 \sum_{\alpha > \beta} {\tilde H}^{\alpha \beta}
	\bbox{\nabla}^2 {\tilde M}_0^{\alpha \beta} \nonumber \\
	&&\qquad + g_0 \sum_{\alpha,\beta} \Bigl( 
	{\tilde h}^\alpha {\tilde M}_0^{\alpha \beta} S_0^\beta - 
	{\tilde H}^{\alpha \beta} {\tilde S}_0^\alpha S_0^\beta -
							\nonumber \\
	&&\qquad \qquad \qquad - \sum_\gamma {\tilde H}^{\alpha \beta}
	{\tilde M}_0^{\alpha \gamma} M_0^{\beta \gamma} \Bigr) 
							\Biggr] \ .
 \label{extjdd}
\end{eqnarray}
Therefore the dynamic order parameter susceptibility becomes
\begin{eqnarray}
	&&\chi_0({\bf x}-{\bf x'},t-t') \delta^{\alpha \beta} =
	{\delta \langle S_0^\alpha({\bf x},t) \rangle \over
		\delta {\tilde h^\beta({\bf x}',t')}}
		\bigg\vert_{{\tilde h}^\beta = 0} \nonumber \\
	&&\quad = \lambda_0 \langle S_0^\alpha({\bf x},t) 
	{\tilde S}_0^\beta({\bf x}',t') \rangle + \nonumber \\
	&&\qquad + g_0 \sum_\gamma \langle S_0^\alpha({\bf x},t)
	\left[ {\tilde M}_0^{\beta \gamma} 
			S_0^\gamma \right]({\bf x}',t') \rangle \ ,
 \label{opsusc}
\end{eqnarray}
and similarly the response function for the conserved quantities reads
\begin{eqnarray}
	&&X_0({\bf x}-{\bf x'},t-t') \left( \delta^{\alpha \gamma} 
	\delta^{\beta \delta} - \delta^{\alpha \delta} 
			\delta^{\beta \gamma} \right) = \nonumber \\
	&&\qquad = {\delta \langle M_0^{\alpha \beta}({\bf x},t)
	\rangle \over \delta {\tilde H^{\gamma \delta}({\bf x}',t')}} 
	\bigg\vert_{{\tilde H}^{\gamma \delta} = 0} \nonumber \\ 
	&&\quad = - D_0 \langle M_0^{\alpha \beta}({\bf x},t)
	\bbox{\nabla}^2 {\tilde M}_0^{\gamma \delta}({\bf x}',t') 
						\rangle - \nonumber \\
	&&\qquad - 2 g_0 \langle M_0^{\alpha \beta}({\bf x},t) \left[ 
	{\tilde S}_0^\gamma S_0^\delta \right]({\bf x}',t') \rangle\ . 
 \label{amsusc}
\end{eqnarray}
[Note that $\sum_\rho \langle M_0^{\alpha \beta}({\bf x},t) 
[ {\tilde M}_0^{\gamma \rho} M_0^{\delta \rho} ]({\bf x}',t') \rangle 
= 0$.] Hence we also need cumulants containing composite operators
$Y_0^\alpha = \sum_\beta [ {\tilde M}_0^{\alpha \beta} S_0^\beta ]$ and
$Y_0^{\alpha \beta} = [ {\tilde S}_0^\alpha S_0^\beta ]$, as well as
the corresponding vertex functions, which are related to each other via
\begin{equation}
	G^c_{0 \, S Y}({\bf q},\omega) = 
	- {\Gamma_{0 \, {\tilde S} Y}({\bf q},\omega) \over 
		\Gamma_{0 \, {\tilde S} S}(-{\bf q},-\omega)} \ .
 \label{compop}
\end{equation}
Using Eqs.~(\ref{verfn1}) and (\ref{compop}), we can finally write
\begin{eqnarray}
	\chi_0({\bf q},\omega) &&= 
	\Gamma_{0 \, {\tilde S} S}(-{\bf q},-\omega)^{-1} \times
						\nonumber \\
	&&\qquad \times \left[ \lambda_0 - g_0 
	\Gamma_{0 \, {\tilde S} [{\tilde M} S]}({\bf q},\omega)
							\right] \ ,
 \label{opvsus} \\
	X_0({\bf q},\omega) &&=
	\Gamma_{0 \, {\tilde M} M}(-{\bf q},-\omega)^{-1} \times
						\nonumber \\
	&&\qquad \times \left[ D_0 q^2 + 2 g_0 
	\Gamma_{0 \, {\tilde M} [{\tilde S} S]}({\bf q},\omega) 
							\right] \ .
 \label{amvsus}
\end{eqnarray}

We conclude this discussion of general properties of the SSS model
with the derivation of Ward identies which are a consequence of the
$O(n)$ symmetry, and the fact that the $M_0^{\alpha \beta}$ are the
generators of this symmetry group \cite{modele,sssfth,uwedis}. As a
first version, consider that a spatially homogeneous, but
time--dependent external field ${\tilde H}^{\alpha \beta}(t)$ is 
switched on at $t = 0$. According to Eq.~(\ref{extfld}) and the
equation of motion (\ref{soplan}), this produces the following
additional contribution to the expectation value of the order
parameter component $S_0^\alpha$,
\begin{equation}
	\langle S_0^\alpha({\bf x},t) \rangle_{\tilde H} = 
	- g_0 \int_0^t \! dt' {\tilde H}^{\alpha \beta}(t')
	\langle S_0^\beta({\bf x},t') \rangle_{\tilde H} \ .
 \label{hopexp}
\end{equation}
Upon employing suitable variational derivatives, this leads to the
following relation between the nonlinear susceptibility
\begin{eqnarray}
	&&R_{0 \, S; S M}({\bf x},t;{\bf x}',t';{\bf x}'',t'')
	\left( \delta^{\alpha \gamma} \delta^{\beta \delta} - 
		\delta^{\alpha \delta} \delta^{\beta \gamma} \right) =
							\nonumber \\ 
	&&\quad = {\delta^2 \langle S_0^\alpha({\bf x},t) \rangle
	\over \delta {\tilde h}^\beta({\bf x}',t') 
		\delta {\tilde H}^{\gamma \delta}({\bf x}'',t'')}
	\bigg\vert_{{\tilde h}^\beta = {\tilde H}^{\gamma \delta} = 0}
 \label{nlsusc}
\end{eqnarray}
and the order parameter response,
\begin{equation}
	\int \! d^dx' R_{0 \, S; S M}({\bf x},t;{\bf 0},0;{\bf x}',t')
			= - g_0 \Theta(t - t') \chi_0({\bf x},t) \ .
 \label{wiresp}
\end{equation}
An equivalent Ward identity for vertex functions can be obtained by
noting that the ``mixed'' generating functional 
$W[\{ {\tilde \phi}_0^\alpha \} , \{ \phi_0^\alpha \} ,
\{ {\tilde H}^{\alpha \beta} \} , \{ H^{\alpha \beta} \}]$, [compare
Eqs.~(\ref{genrfn}), (\ref{vertfn})] is invariant with respect to the
following nontrivial variations corresponding to Eq.~(\ref{hopexp}):
$\delta {\tilde H}^{\alpha \beta} = \varepsilon 
{\tilde H}^{\alpha \beta}$, $\delta \phi_0^\alpha = - \varepsilon g_0
\sum_\beta {\tilde H}^{\alpha \beta} \phi_0^\beta t$ \cite{uwedis}. 
Hence 
\begin{equation}
	\delta W = {\varepsilon \over 2} \! \int \! d^dx \int \! dt 
	\sum_{\alpha , \beta} {\tilde H}^{\alpha \beta} \left[
	{\delta W \over \delta {\tilde H}^{\alpha \beta}} - 2 g_0 
		{\delta W \over \delta \phi_0^\alpha} \phi_0^\beta t 
						\right] = 0 \ ,
 \label{wigenf}
\end{equation}
which with ${\tilde \mu}_0^{\alpha \beta} = \delta \ln Z / \delta 
{\tilde H}^{\alpha \beta}$ translates to a Ward identity for the
generating functional (\ref{vertfn}) of the vertex functions,
\begin{equation}
	\int \! d^dx \int \! dt \sum_{\alpha , \beta} 
	{\delta \Gamma \over \delta {\tilde \mu}_0^{\alpha \beta}}
	\left[ {\tilde \mu}_0^{\alpha \beta} - 2 g_0
	{\delta \Gamma \over \delta \phi_0^\alpha} \phi_0^\beta t 
						\right] = 0 \ .
 \label{wiverf}
\end{equation}
Specifically, this yields
\begin{eqnarray}
	&&\Gamma_{0 \, {\tilde M} {\tilde S} S}({\bf q}/2 , \omega/2 ;
	{\bf q}/2 , \omega/2 ; -{\bf q} , -\omega) = \nonumber \\
	&&= g_0 {\partial \over \partial (i \omega)} \left[ 
	\Gamma_{0 \, {\tilde M} {\tilde M}}({\bf q}/2,\omega/2)
	\Gamma_{0 \, {\tilde S} S}({\bf q}/2,\omega/2) \right] \ .
 \label{widver}
\end{eqnarray}
Note that Eqs.~(\ref{wiresp}) and (\ref{widver}) hold quite
independently of any detailed--balance condition.

We can use these Ward identities now to demonstrate that the
mode--coupling constant $g$, as a consequence of the underlying $O(n)$
symmetry, does not renormalize \cite{sssfth}. First, we note that the
{\em static} response function for the conserved angular momenta is
{\em exactly}
\begin{equation}
	X_0({\bf q},\omega = 0) \equiv 1 \ ,
 \label{stasam}
\end{equation}
as follows from the hamiltonian (\ref{stasss}) and the fact that in
the limit $\omega \to 0$ there is no coupling between the critical
fluctuations $S_0^\alpha$ and $M_0^{\alpha \beta}$, which is true even
for our nonequilibrium model. Therefore there cannot be any field
renormalization for the angular momenta: 
$M^{\alpha \beta} = Z_M^{1/2} M_0^{\alpha \beta}$ (and similarly for
${\tilde M}^{\alpha \beta}$) with $Z_M \equiv 1$. Second, as a result
of the $q$ dependence of the mode--coupling vertices (\ref{mctfun}),
to {\em all orders} in perturbation theory 
\begin{equation}
	{\partial \over \partial (i \omega)} 
	\Gamma_{0\,{\tilde M} M}({\bf q} = {\bf 0},\omega)\equiv 1 \ ,
 \label{amverf}
\end{equation}
and hence $Z_{\tilde M} Z_M \equiv 1$. We remark that an analogous
equation for model B leads to the identity $z = 4 - \eta$ for the
dynamic exponent \cite{bajawa}; a similar result for the KPZ equation
implies the absence of field renormalizations there as well
\cite{kpzfth}. At last, we utilize the above Ward identities
(\ref{wiresp}), (\ref{widver}), both of which imply that the
renormalization factor for the mode--coupling constant is identical to
$Z_g \equiv Z_M \equiv 1$. Physically, this means that the reversible
mode couplings are not affected by critical fluctuations. This fact
will lead to certain general identities for the dynamic exponent in
the scaling regimes, see Sec.~\ref{fpoint}. Again, similar Ward
identities may be derived and corresponding conclusions can be drawn
for the mode--coupling constant in model J, as mentioned above,
leading to the exact result $z = (d + 2 - \eta) / 2$ \cite{modelj},
and also for the nonlinearity in the KPZ problem, there originating in
the Galilean invariance of the equivalent Burgers equation, and
implying the nontrivial scaling relation $z + \chi = 2$ between the
dynamic and roughness exponents \cite{kpzfth}.

\subsection{Renormalization to one--loop order}
 \label{looprg}

Bearing the results of the previous subsection in mind, we introduce
multiplicatively renormalized fields and parameters according to
\begin{eqnarray}
	&&{\tilde S}^\alpha = Z_{\tilde S}^{1/2} {\tilde S}_0^\alpha
	\; , \quad S^\alpha = Z_S^{1/2} S_0^\alpha \ ,
 \label{fldren} \\
	&&{\tilde \lambda} = 
	Z_{\tilde S}^{-1} Z_{\tilde \lambda} {\tilde \lambda}_0 \; , 
	\quad {\tilde D} = Z_{\tilde D} {\tilde D}_0 \ ,
 \label{noiren} \\
	&&\lambda = (Z_{\tilde S} Z_S)^{-1/2} Z_\lambda \lambda_0 \; ,
	\quad D = Z_D D_0 \ , 
 \label{onsren} \\
	&&\tau = Z_S^{-1} Z_\tau \tau_0 \mu^{-2} \; , \quad 
		\tau_0 = r_0 - r_{0c} \ ,
 \label{tauren} \\
	&&u = Z_S^{-2} Z_u u_0 A_d \mu^{d-4} \ . 
 \label{sturen}
\end{eqnarray}
Here, $A_d = \Gamma(3 - d/2) / 2^{d-1} \pi^{d/2}$ denotes a
$d$--dependent geometric factor, and $\mu$ is a momentum scale. Thus
all the renormalized couplings are dimensionless, as is 
$g = g_0 A_d^{1/2} \mu^{(d-4)/2}$. Note that both the static and the
mode--coupling constants $u_0$ and $g_0$ become dimensionless at the
upper critical dimension $d_c = 4$. We determine the renormalization
constants (Z factors) by demanding that they absorb all the
(ultraviolet) divergences in the corresponding vertex functions to
one--loop order (see the Appendix). We furthermore employ the
dimensional regularization scheme with minimal subtraction in 
$d = 4 - \epsilon$ dimensions, i.e., only include the ultraviolet
poles $\propto 1 / \epsilon$ in the Z factors, along with their
residues in four dimensions (further details on these procedures can
be found in Ref.~\cite{stafth}). In order to avoid the infrared
singularities near the critical point, we take $\tau = 1$ 
($\tau_0 = \mu^2$ to one--loop order) and ${\bf q} = {\bf 0}$, 
$\omega = 0$ as the normalization point. This, of course, follows
closely the renormalization procedure for the equilibrium SSS model
\cite{sssfth} (see also Ref.~\cite{oerjan}).

Using $\Gamma_{{\tilde S} {\tilde S}}({\bf q},\omega) = 
Z_{\tilde S}^{-1} \Gamma_{0\, {\tilde S}{\tilde S}}({\bf q},\omega)$,
the renormalization of the noise strengths ${\tilde \lambda}_0$ and
${\tilde D}_0$, as well as of the diffusion constant $D_0$ is readily
inferred from Eqs.~(\ref{g0tsts}), (\ref{g0tmtm}), and (\ref{gm0tmm}),
respectively, with the results
\begin{eqnarray}
	Z_{\tilde \lambda} &&= 
	1 + {A_d \mu^{-\epsilon} \over \epsilon} \, 
			{n - 1 \over 1 + w_0} \, w_0 \, {\bar f}_0 \ ,
 \label{zopnoi} \\
	Z_{\tilde D} &&= 1 + {A_d \mu^{-\epsilon} \over 2 \epsilon} \,
				w_0 \, {\bar f}_0 \, \Theta_0^2 \ ,
 \label{zamnoi} \\
	Z_D &&= 1 + {A_d \mu^{-\epsilon} \over 2 \epsilon} \, 
				w_0 \, {\bar f}_0 \, \Theta_0 \ ,
 \label{zamdif}
\end{eqnarray}
where we have used the definitions (\ref{tratio}) and
\begin{equation}
	w_0 = {\lambda_0  \over D_0} \; , \quad
	{\bar f}_0 = g_0^2 {{\tilde D}_0 \over \lambda_0^2 D_0}
 \label{effcpl}
\end{equation}
for the ratio of relaxation constants $w_0$ and effective dynamical
coupling ${\bar f}_0$.

Next, we consider $\Gamma_{{\tilde S} S}({\bf q},\omega) = 
(Z_{\tilde S} Z_S)^{-1/2} \Gamma_{0 \, {\tilde S} S}({\bf q},\omega)$,
see Eq.~(\ref{gamtss}). First, we determine the fluctuation--induced
$T_c$ shift $r_{0c}$ from the condition of criticality
$\chi_0({\bf q} = {\bf 0},\omega = 0)^{-1} = 0$, which because of
Eq.~(\ref{opvsus}) is equivalent to demanding that 
$\Gamma_{0 \, {\tilde S} S}({\bf 0},0) = 0$ for $r_0 = r_{0c}$. 
Eq.~(\ref{gm0tss}) then yields with Eq.~(\ref{stscal})
\begin{eqnarray}
	&&r_{0c} = - {n + 2 \over 6} \, {\tilde u}_0 \int_k \! 
			{1 \over r_{0c} + k^2} - \nonumber \\ 
	&&- (n - 1) \, w_0 \, {\bar f}_0 \, (1 - \Theta_0)
	\int_k \! {1 \over w_0 \, r_{0c} + (1 + w_0) k^2} \ ;
 \label{tcshif}
\end{eqnarray}
note that for $d \leq 2$ the integrals on the right--hand--side of
Eq.~(\ref{tcshif}) are infrared--divergent, which indicates that
$d_{lc} = 2$ is the lower critical dimension. Evaluating the momentum
integrals for $2 < d < 4$ gives explicitly 
\begin{eqnarray}
	|r_{0c}| = \Biggl( &&{2 A_d \over (d - 2)  (4 - d)} \Biggl[ 
	{n + 2 \over 6} {\tilde u}_0 + (n - 1) \times \nonumber \\
	&&\times \left( {w_0 \over 1 + w_0} \right)^{d/2} 
	{\bar f}_0 \, (1 - \Theta_0) \Biggr] \Biggr)^{2 / (4 - d)}
 \label{r0cexp}
\end{eqnarray}
(notice the pole at $d_{lc} = 2$ and the essential singularity at
$d_c = 4$). The first term here corresponds to the downwards shift of
the critical temperature of the $\bbox{\phi}^4$ model; the second 
contribution, which is of purely dynamical origin, may either reduce
$T_c$ further, namely for $T_S < T_M$, or enhance it with respect to
the equilibrium situation, if $T_S > T_M$.

Upon defining $\tau_0 = r_0 - r_{0c}$, the true distance from the
critical point, and inserting Eq.~(\ref{tcshif}) into (\ref{gm0tss}), 
setting $r_{0c} = 0 + {\cal O}(u_0,g_0^2)$ in the integrals, one finds
\begin{eqnarray}
	Z_\tau {Z_\lambda \over Z_S} = 1 &&- 
	{A_d \mu^{-\epsilon} \over \epsilon} \, {n + 2 \over 6} \,
	{\tilde u}_0 + {A_d \mu^{-\epsilon} \over \epsilon} \,
	{n - 1 \over 1 + w_0} \, w_0 \, {\bar f}_0 \, \Theta_0 
							\nonumber \\
	&&- {A_d \mu^{-\epsilon} \over \epsilon} \, {(n - 1) w_0^2 
		\over (1 + w_0)^2} \, {\bar f}_0 \, (1 - \Theta_0) \ .
 \label{zoptau}
\end{eqnarray}
Then, rendering 
$\partial \Gamma_{{\tilde S} S}({\bf 0},\omega) / \partial (i \omega)$
and $\partial \Gamma_{{\tilde S} S}({\bf q},0) / \partial q^2$ finite
gives
\begin{eqnarray}
	&&(Z_{\tilde S} Z_S)^{1/2} = 
	1 - {A_d \mu^{-\epsilon} \over \epsilon} \, {(n - 1) w_0^2 
		\over (1 + w_0)^2} \, {\bar f}_0 \, (1 - \Theta_0) \ ,
 \label{zfprod} \\
	&&Z_\lambda = 1 + {A_d \mu^{-\epsilon} \over \epsilon} \,
	{n - 1 \over 1 + w_0} \, w_0 \, {\bar f}_0 \, \Theta_0 -
				\nonumber \\ &&\qquad \qquad 
	- {A_d \mu^{-\epsilon} \over \epsilon} \, {(n - 1) w_0^2 \over
		(1 + w_0)^3} \, {\bar f}_0 \, (1 - \Theta_0) \ .
 \label{zoprel}
\end{eqnarray}
Eq.~(\ref{zfprod}) also absorbes the divergences in the three--point
function (\ref{g0tssm}), which confirms that indeed $Z_g = 1$. 
Eq.~(\ref{g0tmss}) can then be used to determine the still unknown
field renormalization itself, with the result
\begin{equation}
	Z_S = 1 - {A_d \mu^{-\epsilon} \over 2 \epsilon} \, {n - 1 
	\over (1 + w_0)^2} \, w_0 \, {\bar f}_0 \, (1 - \Theta_0) \ ;
 \label{zfield}
\end{equation}
notice that $Z_S \not= 1$ and $Z_{\tilde S} \not= 1$ already to
one--loop order if $\Theta_0 \not= 1$. At last, the multiplicative
renormalization of the nonequilibrium SSS model vertex functions is
concluded by rendering the four--point function (\ref{g0tsss}) finite
with
\begin{eqnarray}
	Z_u {Z_\lambda \over Z_S} &&= 1 - 
	{A_d \mu^{-\epsilon} \over \epsilon} \, {n + 8 \over 6} \,
	{\tilde u}_0 + {A_d \mu^{-\epsilon} \over \epsilon} \,
	{n - 1 \over 1 + w_0} \, w_0 \, {\bar f}_0 - \nonumber \\
	&&\qquad - {A_d \mu^{-\epsilon} \over \epsilon} \, 
	{(n - 1) w_0^2 \over (1 + w_0)^2} \, {\bar f}_0 \, 
					(1 - \Theta_0) - \nonumber \\
	&&- {6 A_d \mu^{-\epsilon} \over \epsilon} \,
	{n - 1 \over 1 + w_0} \, {(w_0 {\bar f}_0)^2 \over 
		{\tilde u}_0} \, \Theta_0 \, (1 - \Theta_0) \ .
 \label{zustat}
\end{eqnarray}
When detailed balance holds, $\Theta_0 = 1$, these one--loop Z factors
reduce to the well--known equilibrium results \cite{sssfth}.

Whereas the vertex functions and hence also the two--point correlation
functions (\ref{verfn2}) are rendered finite with the above Z factors,
the dynamic response functions may require additional additive
renormalizations, as a consequence of the involved composite
operators. The response function for the conserved angular momenta
(\ref{amvsus}), using Eq.~(\ref{stasam}) for its static limit, can
generally be written in the following form,
\begin{equation}
	X_0({\bf q},\omega) = {\Delta_0({\bf q},\omega) q^2 \over 
			- i \omega + \Delta_0({\bf q},\omega) q^2} \ .
 \label{amresp}
\end{equation}
Using Eqs.~(\ref{gamtmm}) and (\ref{gtmtss}), one finds the one--loop
result 
\begin{eqnarray}
	&&\Delta_0({\bf q},\omega) = D_0 
	\Biggl[ 1 + {2 \over d} \, w_0 {\bar f}_0 \Theta_0 \int_k \! 
	{k^2 \over \tau_0 + ({\bf q}/2 + {\bf k})^2} \times \nonumber
	\\ &&\times {1 \over \tau_0 + ({\bf q}/2 - {\bf k})^2}
	\, {1 \over -i \omega / 2 \lambda_0 + \tau_0 + q^2 / 4 + k^2}
							\Biggr] \ ;
 \label{amonco}
\end{eqnarray}
hence, as the ultraviolet singularity in (\ref{amonco}) is absorbed by
the Z factor (\ref{zamdif}), no additive renormalization is needed. 
This comes as no surprise, as the contribution from Eq.~(\ref{gtmtss})
is nonsingular. 

However, the integral Eq.~(\ref{gtstms}) is divergent, and therefore a
corresponding additive renormalization has to be introduced. The
structure of the order parameter susceptibility (\ref{opvsus}) is
\begin{equation}
	\chi_0({\bf q},\omega) = {\Lambda_0({\bf q},\omega) \over 
	-i \omega + \Lambda_0({\bf q},\omega) / \chi_0({\bf q},0)} \ . 
 \label{opresp}
\end{equation}
Here, using Eqs.~(\ref{gamtss}) and (\ref{gtstms}), the static
susceptibility reads to one--loop order
\begin{eqnarray}
	&&\chi_0({\bf q},0)^{-1} = 
	\tau_0 \Biggl[ 1 - {n + 2 \over 6} \, {\tilde u}_0 
		\int_k \! {1 \over k^2 (\tau_0 + k^2)} - \nonumber \\
	&&\qquad - {n - 1 \over 1 + w_0} w_0^2 {\bar f}_0 
	(1 - \Theta_0) \times \nonumber \\ &&\quad \times \int_k \! 
	{1 \over k^2 [w_0 \tau_0 - (1 - w_0) ({\bf q} \cdot {\bf k}) +
				(1 + w_0) (q^2/4 + k^2)]} \nonumber \\ 
	&&\quad + q^2 + {n - 1 \over 1+ w_0} w_0 {\bar f}_0 
	(1 - \Theta_0) \times \nonumber \\ &&\times \int_k \! 
	{(1 - w_0) ({\bf q} \cdot {\bf k}) - (1 + w_0) q^2/4 \over 
	k^2 [w_0 \tau_0 - (1 - w_0) ({\bf q} \cdot {\bf k})
		+ (1 + w_0) (q^2/4 + k^2)]} \Biggr] \ , \nonumber \\
 \label{opsres}
\end{eqnarray}
where Eq.~(\ref{tcshif}) has been inserted, and the renormalized
Onsager coefficient is 
\begin{eqnarray}
	&&\Lambda_0({\bf q},\omega) = \lambda_0 
	\Biggl[ 1 + (n - 1) w_0 {\bar f}_0 \Theta_0 \int_k \!
	{1 \over \tau_0 + ({\bf q}/2 + {\bf k})^2} \times \nonumber \\
	&&\times {1 \over - i \omega / D_0 + w_0 [\tau_0 + ({\bf q}/2 
		+ {\bf k})^2] + ({\bf q}/2 - {\bf k})^2} \Biggr] \, ;
 \label{oponco}
\end{eqnarray}
as expected, the above multiplicative renormalizations with
Eqs.~(\ref{zoptau})--(\ref{zfield}) do {\em not} suffice to remove the
divergences in Eqs.~(\ref{opsres}) and (\ref{oponco}). We determine
the necessary additive renormalization by requiring that 
\begin{equation}
	{\partial \over \partial q^2} 
		\chi_0({\bf q},0)^{-1}_{\rm sing} = Z_S + A_S \ ,
 \label{addren}
\end{equation}
and Eqs.~(\ref{opsres}) and (\ref{zfield}) then yield
\begin{equation}
	A_S = - {A_d \mu^{-\epsilon} \over \epsilon} \,
	{n - 1 \over (1 + w_0)^2} \, \left( {1 \over 2} + 
	{w_0 \over 1 + w_0} \right) w_0 {\bar f}_0 (1 - \Theta_0) \ .
 \label{addsin}
\end{equation}
Indeed, $\chi_0({\bf 0},0)^{-1}$ and $\Lambda_0({\bf 0},0)$ are then
rendered finite with the combinations of Z factors 
$(Z_S + A_S) Z_\tau / Z_S$ and $Z_\lambda / (Z_S + A_S)$,
respectively.

\subsection{RG flow equations and fixed points}
 \label{fpoint}

The renormalization group equations serve to connect the asymptotic
theory, where the infrared divergences become manifest, with a region
in parameter space (in our case consisting of $\{ a \} = 
{\tilde \lambda} , {\tilde D} , \lambda , D , g , u , \tau$) where the
couplings are finite and ordinary ``naive'' perturbation expansion is
applicable. They are derived by observing that the ``bare'' vertex
functions do not depend on the renormalization scale $\mu$,
\begin{equation}
	\mu {d \over d\mu} \bigg\vert_0 \Gamma_{0 \, 
	{\tilde S}^r {\tilde M}^k S^s M^l}(\{ a_0 \}) = 0 \ .
 \label{rngreq}
\end{equation}
Introducing Wilson's flow functions
\begin{eqnarray}
	\zeta_{\tilde S} &&= \mu {\partial \over \partial \mu}
		\bigg\vert_0 \ln Z_{\tilde S} \; , \quad
	\zeta_S = \mu {\partial \over \partial \mu} 
		\bigg\vert_0 \ln Z_S \ ,
 \label{zetfld} \\
	\zeta_a &&= \mu {\partial \over \partial \mu}
		\bigg\vert_0 \ln {a \over a_0} \ .
 \label{zetpar}
\end{eqnarray}
Eq.~(\ref{rngreq}) may be written as a partial differential equation
for the renormalized vertex functions
\begin{eqnarray}
	&&\left[ \mu {\partial \over \partial \mu} + \! \sum_{\{ a \}}
	\zeta_a a {\partial \over \partial a} + {r \over 2}
	\zeta_{\tilde S} + {s \over 2} \zeta_S \right] \times
	\nonumber \\ &&\qquad \qquad \qquad \times 
	\Gamma_{{\tilde S}^r {\tilde M}^k S^s M^l}(\mu,\{ a \})= 0 \ .
 \label{calsym}
\end{eqnarray}
Note that $\zeta_{\tilde M} = \zeta_M \equiv 0$ and 
$\zeta_g \equiv - \epsilon / 2$ as a consequence of the exact results
in Sec.~\ref{genres}. Eq.~(\ref{calsym}) can be solved with the method
of characteristics $\mu \to \mu \ell$; this defines running couplings
as the solutions to the first--order differential RG flow equations
\begin{equation}
	\ell {d a(\ell) \over d\ell} = \zeta_a(\ell) a(\ell) \; ,
	\quad a(1) = a \ .	
 \label{rgflow}
\end{equation}
The solution of the Callan--Symanzik equation (\ref{calsym}) then
reads 
\begin{eqnarray}
	&&\Gamma_{{\tilde S}^r {\tilde M}^k S^s M^l}
		(\mu,\{ a \},q,\omega) = \nonumber \\
	&&\quad \exp \left\{ {1 \over 2} \int_1^\ell \! 
	\Bigl[ r \zeta_{\tilde S}(\ell') + s \zeta_S(\ell') \Bigr] 
		{d \ell' \over \ell'} \right\} \times \nonumber \\ 
	&&\; \times \Gamma_{{\tilde S}^r {\tilde M}^k S^s M^l}
	(\mu \ell,\{ a(\ell) \},q/\mu\ell,\omega/\mu^2\ell^2) \ . 
 \label{solcsy}
\end{eqnarray}

Upon introducing the renormalized ratios
\begin{equation}
	w = {\lambda \over D} \; , \quad \Theta = 
	{{\tilde \lambda} \over \lambda} \, {D \over {\tilde D}} 
 \label{rratio}
\end{equation}
and renormalized effective couplings
\begin{equation}
	{\bar f} = g^2 \, {{\tilde D} \over \lambda^2 D} \; , \quad
	{\tilde u} = {{\tilde \lambda} \over \lambda} \, u \ ,
 \label{rcoupl}
\end{equation}
and collecting the definitions Eqs.~(\ref{fldren})--(\ref{tauren}) and
one--loop results (\ref{zopnoi})--(\ref{zamdif}) and
(\ref{zoptau})--(\ref{zfield}), one finds
\begin{eqnarray}
	\zeta_S &=& {n - 1 \over 2} \, {1 \over (1 + w)^2} \,
				w \, {\bar f} \, (1 - \Theta) \ ,
 \label{zetafd} \\
	\zeta_{\tilde S} &=& - {n - 1 \over 2} \, 
	{1 - 4 w \over (1 + w)^2} \, w \, {\bar f} \, (1 - \Theta) \ ,
 \label{zettfd} \\
	\zeta_\tau &=& - 2 + {n + 2 \over 6} \, {\tilde u} + {(n - 1) 
	w^2 \over (1 + w)^3} \, w \, {\bar f} \, (1 - \Theta) \ ,
 \label{zettau} \\
 	\zeta_{\tilde \lambda} &=& - {n - 1 \over 1 + w} \, 
	w \, {\bar f} + {n - 1 \over 2} \, {1 - 4 w \over (1 + w)^2}
			\, w \, {\bar f} \, (1 - \Theta) \ , 
 \label{zetopn} \\
	\zeta_{\tilde D} &=& 
		- {1 \over 2} \, w \, {\bar f} \, \Theta^2 \ ,
 \label{zetamn} \\
	\zeta_{\lambda} &=& - {n - 1 \over 1 + w} \, 
	w \, {\bar f} \, \Theta - {(n - 1) w^2 \over (1 + w)^3} \, 
				w \, {\bar f} \, (1 - \Theta) \ ,
 \label{zetopr} \\
	\zeta_D &=& - {1 \over 2} \, w \, {\bar f} \, \Theta \ .
 \label{zetamd}
\end{eqnarray}
Notice that nonzero values of $\zeta_{\tilde \lambda}$ and
$\zeta_{\tilde D}$ induce anomalous noise correlations, while
$\zeta_\lambda$ and $\zeta_D$ determine the dynamic critical
exponents, see Eq.~(\ref{dynexp}) below.

We furthermore need the flows for the running couplings $v(\ell)$,
with $\{ v \} = w , \Theta , {\bar f}, {\tilde u}$,
\begin{equation}
	\ell {d v(\ell) \over d\ell} = \beta_v(\ell) \; , 
	\quad v(1) = v \ ,
 \label{runcpl}
\end{equation}
as given by the beta functions
\begin{equation}
	\beta_v = 
	\mu {\partial \over \partial \mu} \bigg\vert_0 \, v \ ;
 \label{betafn}
\end{equation}
with Eqs.~(\ref{sturen}), (\ref{zustat}), and
(\ref{zetopn})--(\ref{zetamd}) these become to one--loop order
\begin{eqnarray}
	\beta_w &&= w \left( \zeta_\lambda - \zeta_D \right) \nonumber
	\\ &&= w^2 \, {\bar f} \left[ \left( {1 \over 2} - 
	{n - 1 \over 1 + w} \right) \Theta - {(n - 1) w^2 \over 
	(1 + w)^3} \, (1 - \Theta) \right] \ , \nonumber \\
 \label{betaww} \\
	\beta_\Theta &&= \Theta \left( \zeta_{\tilde \lambda} -
	\zeta_{\tilde D} - \zeta_\lambda + \zeta_D \right) \nonumber
	\\ &&= - {1 \over 2} \, w \, {\bar f} \, \Theta (1 - \Theta)
	\left[ \Theta + (n - 1) {1 + 7 w + 4 w^2 \over (1 + w)^3} 
					\right] \ , \nonumber \\
 \label{betath} \\
	\beta_{\bar f} &&= {\bar f} \left( - \epsilon + 
	\zeta_{\tilde D} - 2 \zeta_\lambda - \zeta_D \right) \nonumber
	\\ &&= {\bar f} \Biggl[ - \epsilon - 
	{1 \over 2} \, w \, {\bar f} \, \Theta^2 + 
	\left( {1 \over 2} + {2 (n - 1) \over 1 + w} \right) 
			w \, {\bar f} \, \Theta + \nonumber \\
	&&\qquad + {2 (n - 1) w^2 \over (1 + w)^3} \, 
			w \, {\bar f} \, (1 - \Theta) \Biggr] \ ,
 \label{betafb} \\
	\beta_{\tilde u} &&= {\tilde u} \Biggl[ - \epsilon + 
	{n + 8 \over 6} \, {\tilde u} - {2 (n - 1) w \over (1 + w)^3}
		\, w \, {\bar f} \, (1 - \Theta) - \nonumber \\
	&&\quad - {2 (n - 1) \over 1 + w} 
	\left( 1 - {3 w {\bar f} \Theta	\over {\tilde u}} \right) 
			w \, {\bar f} \, (1 - \Theta) \Biggr] \ . 
 \label{betatu}
\end{eqnarray}

We are now ready to explore the fixed points of the RG flow equations,
as given by the zeros of the beta functions
(\ref{betaww})--(\ref{betatu}). First, we can check that indeed for 
$\Theta^* = 1$ the equilibrium fixed points (see Ref.~\cite{sssfth})
emerge. The above flow equations then simplify considerably, and the
effective dynamical coupling in Eqs.~(\ref{zetopn})--(\ref{zetamd})
becomes
\begin{equation}
	f = w \, {\bar f} = {g^2 \over \lambda D} \ ,
 \label{eqeffc}
\end{equation}
because as now $\zeta_{\tilde \lambda} \equiv \zeta_\lambda$ and
$\zeta_{\tilde D} \equiv \zeta_D$, we can identify the noise strenghts
${\tilde \lambda}$ and ${\tilde D}$ with the Onsager coefficients
$\lambda$ and $D$, respectively. The corresponding beta function for
$f$ reads
\begin{equation}
	\beta_f = f (- \epsilon - \zeta_\lambda - \zeta_D) =
	f \left[ - \epsilon + 
	\left( {1 \over 2} + {n - 1 \over 1 + w} \right) f \right] \ .
 \label{beteqf}
\end{equation}
The first equation in (\ref{beteqf}) implies that for {\em any}
nontrivial fixed point $0 < f^* < \infty$ the {\em exact} relation 
$\zeta_\lambda^* + \zeta_D^* = - \epsilon = d - 4$ holds. Furthermore
the analysis of the RG equation in the vicinity of $f^*$ reveals that
the dynamic exponents for the fluctuations of the order parameter and
conserved quantities are given by 
\begin{equation}
	z_S = 2 + \zeta_\lambda^* \; , \quad
	z_M = 2 + \zeta_D^* \ ,
 \label{dynexp}
\end{equation}
which then leads to the following identity \cite{modele,uwedis}
\begin{equation}
	z_s + z_M = d \ .
 \label{expoid}
\end{equation}
Therefore, in a {\em strong--scaling} situation where the
characteristic time scales for the order parameter and angular momenta
are the same, $0 < w^* < \infty$, and hence $z_S = z_M = z$, one finds
the well--known {\em exact} result
\begin{equation}
	z = d / 2 \ .
 \label{strscz}
\end{equation}
Indeed, the above one--loop flow equations (\ref{betaww}) and
(\ref{beteqf}) provide the strong--scaling fixed point
\begin{equation}
	w_{\rm eq}^* = 2 n - 3 \; , \quad f_{\rm eq}^* = \epsilon \ .
 \label{stscfp}
\end{equation}

However, in addition there are two nontrivial {\em weak--scaling}
fixed points with $\zeta_\lambda^* \not= \zeta_D^*$, namely
\begin{equation}
	w_w^* = 0 \; , \quad f_w^* = {2 \epsilon \over 2 n - 1} \ ,
 \label{wscfp1}
\end{equation}
with
\begin{eqnarray}
	z_S &&= 2 - {2 (n - 1) \epsilon \over 2 n - 1}
			+ {\cal O}(\epsilon^2) \ , \nonumber \\
	z_M &&= 2 - {\epsilon \over 2 n - 1} + {\cal O}(\epsilon^2)\ ,
 \label{wkscz1}
\end{eqnarray}
and
\begin{equation}
	w_{w'}^* = \infty \; , \quad f_{w'}^* = 2 \epsilon \ ,
 \label{wscfp2}
\end{equation}
implying that
\begin{equation}
	z_S = 2 \; , \quad z_M = d - 2 \ .
 \label{wkscz2}
\end{equation}
Note that at both fixed points the relation $z_S + z_M = d$ holds, of
course. For the fixed point (\ref{wscfp2}) this sum rule actually even
implies that (\ref{wkscz2}) is probably exact, for $\zeta_\lambda$
should vanish if $w^* = \infty$. Finally, there is also the (Gaussian)
model A fixed point with $f_0^* = 0$ (and $w^*$ unspecified because
now the order parameter and angular momenta are decoupled), with 
$z_S = 2 + {\cal O}(\epsilon^2)$; yet, according to
Eq.~(\ref{beteqf}), it is clearly unstable against $f$ for $d < 4$:
$\ell d f / d \ell = \beta_f(\ell) = - \epsilon f$ at $f_0^* = 0$, and
hence $f$ will increase in the asymptotic limit $\ell \to 0$. 
Similarly, both weak--scaling fixed points are unstable (to one--loop
order at least) for $d < 4$: Near $w_{w'}^* = \infty$ one has 
$\beta_w = \epsilon w$, and hence $w$ will decrease as $\ell \to 0$,
while in the vicinity of $w_w^* = 0$ one finds 
$\beta_w = - \epsilon w (2 n - 3) / (2 n - 1)$, and upon decreasing
$\ell$, $w$ will go down as well. Stability analysis therefore
demonstrates that to one--loop order the {\em strong--scaling} fixed
point with $z = d / 2$ is {\em stable} \cite{modele,modelg,sssfth}; 
however, as $w_w^* = 0$ is actually close to its stability boundary
for $n = 2$, it may well be that to higher loop orders the
strong--scaling fixed point actually becomes unstable for the planar
model, and in fact (\ref{strscz}) does not hold \cite{sssfth,modelf}. 

We shall not pursue this issue further here, but rather turn to the
newly emerging, genuinely {\em nonequilibrium} fixed points, at which
even the static critical behaviour might be changed, as opposed to the
equilibrium situation with $\Theta^* = 1$ for which statics and
dynamics decouple, see Eqs.~(\ref{zetafd}), (\ref{zettau}), and
(\ref{betatu}). For all the above fixed points, we therefore have the
nontrivial static Heisenberg fixed point (to one--loop order)
\begin{equation}
	u_{\rm H}^* = {6 \epsilon \over n + 8} \ ,
 \label{heisen}
\end{equation}
which is stable for $d < 4$, and leads to the critical exponents of
the $O(n)$--symmetric $\bbox{\phi}^4$ model
\begin{eqnarray}
	\eta &=& - \zeta_S^* = 0 + {\cal O}(\epsilon^2) \ , \nonumber \\
	1 / \nu &=& - \zeta_\tau^* = 2 - 
	{(n + 2) \epsilon \over n + 8} +  {\cal O}(\epsilon^2) \ . 
 \label{staexp}
\end{eqnarray}
At the critical point there also appear anomalous long--range noise
correlations, both for the fluctuations of the order parameter and of
the conserved fields. For the order parameter noise, given by
$\Gamma_{{\tilde S}{\tilde S}}({\bf q},\omega)$, one finds 
\begin{equation}
	\langle \zeta^\alpha({\bf q},\omega) 
		\zeta^\beta({\bf q}',\omega') \rangle = 
		\Lambda({\bf q},\omega) \delta({\bf q} + {\bf q}') 
		\delta(\omega + \omega') \delta^{\alpha \beta} \ ,
 \label{eqlopn}
\end{equation}
with 
\begin{equation}
	\Lambda({\bf q},0) \propto q^{z_S - 2} \; , \quad
	\Lambda({\bf 0},\omega) \propto \omega^{(z_S - 2)/2} \ .
 \label{eqopns}
\end{equation}
Similarly, from $\Gamma_{{\tilde M}{\tilde M}}({\bf q},\omega)$ we can
infer the noise for the generalized angular momenta at the critical
point, taking into account its diffusive character,
\begin{eqnarray}
	\langle \eta^{\alpha \beta}({\bf q},\omega) 
		\eta^{\gamma \delta}({\bf q}',\omega') \rangle &&= 
	\Delta({\bf q},\omega) \delta({\bf q} + {\bf q}') 
	\delta(\omega + \omega') \times \nonumber \\
	&&\times \left( \delta^{\alpha \gamma} \delta^{\beta \delta} -
	\delta^{\alpha \delta} \delta^{\beta \gamma} \right) \ ;
 \label{eqlcqn}
\end{eqnarray}
taking into account its diffusive character, we find the limiting
behavior 
\begin{equation}
	\Delta({\bf q},0) \propto q^{z_M} \; , \quad
	\Delta({\bf q} \to {\bf 0},\omega / q^2) \propto 
				(\omega / q^2)^{z_M / 2} \ .
 \label{eqcqns}
\end{equation}
This concludes our discussion of the equilibrium fixed points, and we
now turn to the two new universality classes appearing as a
consequence of our genuinely nonequilibrium perturbation.

The beta function for the temperature ratio $\Theta_0 = T_S / T_M$
(\ref{betath}) reveals that there can only be nonequilibrium fixed
points with either (i) $\Theta^* = 0$ or (ii) $\Theta^* = \infty$,
which as  $T_S \approx T_c$ effectively either correspond to a
``renormalized'' temperature $T_M = \infty$ or $T_M = 0$. In the first
case, $\Theta^* = 0$, one finds 
$\beta_w = - (n - 1) w {\bar f} w^3 / (1 + w)^3$, with the stable
fixed point $w^* = \infty$. Inserting this into Eqs.~(\ref{betafb})
and (\ref{betatu}) yields
\begin{eqnarray}
	\Theta^* = 0 \, : \quad w^* &&= \infty \; , \quad
	{\bar f}^* = {\epsilon \over 2 (n - 1)} \ , \nonumber \\
	{\tilde u}^* &&= 2 u_H^* = {12 \epsilon \over n + 8} \ .
 \label{neqfp1}
\end{eqnarray}
The divergence of $w^*$ already shows that this fixed point describes
a kind of weak--scaling behavior somewhat similar to the equilibrium
fixed point (\ref{wscfp2}) with (\ref{wkscz2}). Indeed, inserting
(\ref{neqfp1}) into Eqs.~(\ref{zetopr}) and (\ref{zetamd}) yields
\begin{equation}
	z_S = d / 2 \; , \quad z_M = 2 \ ,
 \label{neqlz1}
\end{equation}
i.e., the dynamic exponent for the order parameter is identical with
its equilibrium value, while the angular momenta are described by
mean--field theory (simple diffusion). Actually, any nontrivial fixed
point $0 < {\bar f}^* < \infty$ via Eqs.~(\ref{betafb}) and
(\ref{dynexp}) implies the identity $2 z_S + z_M = d + 2 + \zeta_D^*$,
and as for $\Theta^* = 0$ both the anomalous dimensions (\ref{zetamd})
and (\ref{zetamn}) should vanish according to the general structure of
the couplings, the result (\ref{neqlz1}) is probably exact. There is,
however, a nonzero anomalous dimension for the order parameter noise,
$\zeta_{\tilde \lambda}^* = - \epsilon$; therefore the constant 
${\tilde \lambda}_0$ in Eq.~(\ref{sopnoi}) is to be replaced by a
wavevector-- and frequency--dependent function,
\begin{equation}
	\langle \zeta^\alpha({\bf q},\omega) 
		\zeta^\beta({\bf q}',\omega') \rangle = 
	{\tilde \Lambda}({\bf q},\omega) \delta({\bf q} + {\bf q}') 
	\delta(\omega + \omega') \delta^{\alpha \beta} \ ,
 \label{renopn}
\end{equation}
with the singular large--wavelength and low--frequency behavior 
($d < 4$) 
\begin{equation}
	{\tilde \Lambda}({\bf q},0) \propto q^{d - 4} \; , \quad
	{\tilde \Lambda}({\bf 0},\omega) \propto \omega^{(d-4)/2} \ ,
 \label{sinopn}
\end{equation}
which follows from the matching conditions $\mu \ell = q$ and 
$(\mu \ell)^2 = \omega$, respectively. Finally, the new nonequilibrium
static fixed point ${\tilde u}^* = 2 u_H^*$ along with the fact that
now the dynamics affects the static anomalous dimension
(\ref{zettau}), yield the {\em new} static critical exponents
\begin{eqnarray}
	\eta &=& - \zeta_S^* = 0 + {\cal O}(\epsilon^2) \ , \nonumber \\
	1 / \nu &=& - \zeta_\tau^* = 2 - 
	\left( {1 \over 2} + {2 (n + 2) \over n + 8} \right) \epsilon 
					+ {\cal O}(\epsilon^2) \ .
 \label{neqexp}
\end{eqnarray}
According to Eq.~(\ref{dyscal}), the fixed point (\ref{neqfp1})
corresponds to the situation where there exists a coupling 
$\propto g_0$ of the order parameter to the angular momenta, leading
to the dynamic exponent $z_S = d / 2$, the generation of long--range
noise correlations, and even to anomalous static exponents, yet the
dynamics of the conserved quantities themselves remains unaffected by
the critical fluctuations and hence displays mean--field behavior.

The above analysis has tacitly assumed that 
${\tilde w} = w \Theta = {\tilde \lambda} / {\tilde D}$ remains finite
for $\Theta^* = 0$ and $w^* = \infty$. However, 
$\beta_{\tilde w} = {\tilde w} (\zeta_{\tilde \lambda} - 
\zeta_{\tilde D}) = - \epsilon {\tilde w}$, and hence the fixed point
(\ref{neqfp1}) is {\em unstable} for $d < 4$ against an increasing
coupling ${\tilde w}$, or equivalently, against the generation of the
new effective coupling
\begin{equation}
	{\tilde f} = w {\bar f} \Theta^2 
	= g^2 \, {{\tilde \lambda}^2 \over \lambda^3 {\tilde D}} \ ,
 \label{neqefc}
\end{equation}
which characterizes the second nonequilibrium fixed point where 
$\Theta^* = \infty$, and which describes a coupling of the critical
order parameter fluctuations into the diffusive dynamics of the
angular momenta, but no effect of the latter on the equation of motion
for the order parameter itself. Indeed, the anomalous dimensions 
(\ref{zetafd})--(\ref{zetopn}), (\ref{zetopr}), (\ref{zetamd}) all
vanish when $\Theta \to \infty$ with ${\tilde f}$ held finite, and
with Eq.~(\ref{betatu}) one finds the standard $\bbox{\phi}^4$
Heisenberg static exponents (\ref{staexp}) along with model A and
purely diffusive dynamics for the order parameter and conserved
quantities, respectively, 
\begin{equation}
	z_S = 2 + {\cal O}(\epsilon^2) \; , \quad z_M = 2 \ ;
 \label{neqlz2}
\end{equation}
note that $w^*$ cannot be specified because of this decoupling of the
modes. The beta function for the effective mode coupling
(\ref{neqefc}) becomes $\beta_{\tilde f} = {\tilde f} (- \epsilon + 
2 \zeta_{\tilde \lambda} - 3 \zeta_\lambda - \zeta_{\tilde D}) = 
{\tilde f} (- \epsilon + {\tilde f} / 2)$, and hence we arrive at the
following fixed--point values
\begin{equation}
	\Theta^* = \infty \, : \quad {\tilde f}^* = 2 \epsilon  \; ,
	\quad {\tilde u}^* = u_H^* \ .
 \label{neqfp2}
\end{equation}
At this fixed point anomalous noise correlations for the angular
momenta emerge, which in analogy with (\ref{renopn}) can be written in
the form
\begin{eqnarray}
	\langle \eta^{\alpha \beta}({\bf q},\omega) 
		\eta^{\gamma \delta}({\bf q}',\omega') \rangle &&= 
	{\tilde \Delta}({\bf q},\omega) \delta({\bf q} + {\bf q}') 
	\delta(\omega + \omega') \times \nonumber \\
	&&\times \left( \delta^{\alpha \gamma} \delta^{\beta \delta} -
	\delta^{\alpha \delta} \delta^{\beta \gamma} \right) \ ,
 \label{renamn}
\end{eqnarray}
replacing Eq.~(\ref{samnoi}). Their singular behavior follows from
Eq.~(\ref{zetamn}) with (\ref{neqfp2}),
\begin{equation}
	{\tilde \Delta}({\bf q},0) \propto q^\rho \; , \quad 
	\rho = d - 2
 \label{sinamn}
\end{equation}
which is probably an exact result again, because the existence of a
nontrivial fixed point ${\tilde f}^*$ with 
$\rho = 2 + \zeta_{\tilde D}^*$ implies the relation 
$3 z_S + \rho = d + 4 + 2 \zeta_{\tilde \lambda}^*$, and both the
structure of the perturbation theory and the above physical
interpretation require that 
$\zeta_\lambda^* = \zeta_{\tilde \lambda}^* = 0$. Similarly, the
frequency dependence of ${\tilde \Delta}({\bf q},\omega)$ displays
anomalous behavior, but because of the underlying diffusive dynamics
we now have to take the limit ${\bf q} \to {\bf 0}$ more carefully,
namely with $\omega / q^2$ held fixed. One then gets
\begin{equation}
	{\tilde \Delta}({\bf q} \to {\bf 0},\omega / q^2) 
			\propto (\omega / q^2)^{\rho / 2} \ .
 \label{sinanf}
\end{equation}
It is a remarkable fact that these anomalous noise correlations always
appear for those degrees of freedom, {\em towards} which the energy
flows, i.e., those quantitites which are in contact with the heat bath
at {\em lower} temperature. It should be noticed that the power laws
in Eqs.~(\ref{sinopn}) and (\ref{sinamn}), (\ref{sinanf}) are 
{\em not} determined by the corresponding dynamic exponents, as
opposed to the equilibrium situation described by Eqs.~(\ref{eqopns})
and (\ref{eqcqns}), which are consequences of detailed balance. Yet
again, inspection of Eq.~(\ref{betafb}) in the vicinity of
(\ref{neqfp2}), $\beta_{\bar f} = - 2 \epsilon {\bar f}$, shows that
this second new fixed is {\em unstable} with respect to the flow of
${\bar f}$.

Therefore, as both new nonequilibrium fixed points are actually
unstable for $d < 4$, the {\em asymptotic critical behavior} must be
governed by the {\em equilibrium} strong--scaling fixed point
(\ref{stscfp}), (\ref{heisen}) with dynamic exponent (\ref{strscz})
and the usual static critical exponents (\ref{staexp}). The stability
against perturbations away from $\Theta^* = 1$ is readily demonstrated
by observing that $\beta_\Theta = - C (1 - \Theta)$, with $C > 0$
according to Eq.~(\ref{betath}). Hence any deviation from the
equilibrium fixed point will be counteracted by the flow of the
coupling $\Theta$; i.e., whenever the initial value of $\Theta$ is
neither zero or infinite, the flow will asymptotically approach the
stable equilibrium fixed point, and thereby detailed balance is 
{\em dynamically restored}. This is illustrated in Fig.~\ref{flodia},
which displays the location of the fixed points and their flow in the
space of the dynamical couplings $w$, ${\bar f}$, and ${\tilde f}$;
notice that both the equilibrium and ``nonequilibrium'' model A fixed
points are represented by lines here, because the undetermined value
of $w^*$. Concluding this section, we remark that at each of the fixed
points discussed above, including both new nonequilibrium fixed
points, the anomalous dimension stemming from the additive
renormalization (\ref{addsin}) vanishes, as does $\zeta_S$ from the
field renormalization. We therefore did not have to take its effects
into account explicitly.

\section{Summary and conclusions}
 \label{sumcon}

We have studied the critical dynamics of $O(n)$--symmetric systems,
including reversible mode--coupling terms, in the framework of
effective Langevin equations and dynamic field theory, and generalized
the equations of motions to nonequilibrium situations where detailed
balance is broken. In Sec.~\ref{modelj} we have argued that for the
dynamics of isotropic Heisenberg ferromagnets the effect of violating
the Einstein relation between the spin diffusion constant and the
Langevin noise strength can be absorbed via rescaling the static
nonlinearity and the mode--coupling constant. Hence universal
properties cannot be affected by this specific form of
detailed--balance violation, and the critical point is described by
the usual Heisenberg model exponents with the equilibrium dynamic
critical exponent  $z = (d + 2 - \eta) / 2$ \cite{modelj}. This fact
generalizes previous results regarding the stability of the
relaxational models A and B against nonequilibrium perturbations
\cite{grins1,kevbea} to a situation where reversible mode--coupling
terms are present as well.

On the other hand, for the critical dynamics of planar ferromagnets
(model E) \cite{modele} or isotropic antiferromagnets (model G)
\cite{modelg}, both incorporated in the $O(n)$--symmetric SSS model
\cite{sssmod}, such a simple rescaling does not remove the effects of
detailed--balance violation completely, as discussed in
Sec.~\ref{modsss}. This is because there appears a new degree of
freedom, namely the temperature ratio $\Theta_0$ of the heat baths to
which the order parameter and the conserved angular momenta are
attached. This new variable induces different renormalizations for the
noise strengths as compared to the Onsager coefficients
(Sec.~\ref{looprg}), and therefore genuinely new dynamic and static
critical behavior may emerge. Indeed, there appear two new fixed
points of the resulting RG flow equations, describing continuous phase
transitions of entirely nonequilibrium character, namely corresponding
to either $\Theta = 0$ or $\Theta = \infty$, where in the former case
even the static critical exponents become modified. The ensuing dynamic
exponents may be interpreted physically by noting that in both cases
either the coupling of the order parameter into the diffusion equation
for the conserved fields vanishes, while the reverse coupling remains
effective, or vice versa. However, the stability analysis in
Sec.~\ref{fpoint} shows that both of these fixed points are actually
unstable, and provided $0 < \Theta_0 < \infty$ the asymptotic critical
behavior is governed by the nontrivial strong--scaling equilibrium
fixed point of the SSS model characterized again by the static
exponents of the $O(n)$--symmetric Heisenberg model and by the dynamic
exponent $z = d / 2$ \cite{sssmod,sssfth}.

This suggests that the role of detailed balance is actually a weaker
one as compared to internal symmetries; for while breaking a discrete
or continuous symmetry typically results in a change of the
universality class, detailed balance becomes restored here at the
critical point, obviously as a consequence of the underlying $O(n)$
rotation symmetry and the spatial isotropy of the model. We remark
that there are some notable exceptions, though, for the relevance of
symmetry--breaking terms. E.g., the $O(n)$ symmetry of the Heisenberg
model (\ref{hamilt}) is restored as a consequence of the large
critical fluctuations, even when cubic anisotropies are added, if 
$n < n_c$ (with $n_c = 4$ to one--loop order) \cite{cubani}. We should
also cautiously state that our above results relied on the one--loop
approximation only, and specifically the stability boundaries might
change in more accurate calculations, as seems indeed to be the case
for the equilibrium planar model \cite{modelf}. However, in the beta
functions $\beta_{\tilde w}$ and $\beta_{\bar f}$ describing the
instabilities of the fixed points (\ref{neqfp1}) and (\ref{neqfp2}),
respectively, no dangerous $n$--dependences appeared, and it is
therefore probably safe to say that these nonequilibrium fixed points
will remain unstable even to higher orders in perturbation theory,
albeit for $n = 2$ the asymptotic region may ultimately be governed by
the weak--scaling equilibrium fixed point (\ref{wscfp1}), see also
Fig.~\ref{flodia}. 

The remarkable result that violation of detailed balance appears to be
an irrelevant perturbation in the RG sense, at least in
rotation--invariant and isotropic systems, is of course strengthening
the notion of universality even in nonequilibrium situations. Probably
in many experiments probing the critical dynamics with electromagnetic
radiation or neutron scattering, the system is not perfectly
thermalized and variations in the effective temperatures for the
different degrees of freedom cannot be avoided completely due to the
inevitable critical slowing down which prevents fast relaxation
processes. Now if such a perturbation were relevant, its effects would
become enhanced drastically in the vicinity of the phase transition,
however small the initial deviations from equilibrium might have
been. Such a behavior is apparently not observed; however, it might be
interesting to prepare a non--thermalized system on purpose, say by
arranging for the in--plane spins $S^x$ and $S^y$ in a planar
ferromagnet to be on a different temperature than $S^z$, conceivably
attainable with polarized electromagnetic radiation. Another
possibility would be to introduce a long wavelength magnetic field
with random time variation. In an antiferromagnetic material, its
effect would cancel for the staggered magnetization but the coupling
to the magnetization would mimic a situation with $T_M = \infty$.
For both experimental realizations, or at least in equivalent computer
simulations studies, the effects of the above two new fixed points may
then be explored, for these should manifest themselves in nontrivial
crossover behavior, and perhaps even the associated anomalous critical
exponents and noise correlations might be detected. 

We finally remark again that the specific violation of detailed balance
investigated here was {\em isotropic} in character, thereby disturbing
neither of the underlying symmetries of the SSS model, namely $O(n)$
symmetry and its spatial isotropy. Obviously, it would be interesting
to see if the above stability against nonequilibrium perturbations
persists even when the detailed--balance violation is applied in an
anisotropic manner, e.g., by coupling the order parameter to conserved
angular momenta as above, but arranging the noise strength of the
conserved fields to be related to different temperatures in different
space directions. The study of this, or similar direction-- or
scale--dependent nonequilibrium perturbations provides a promising
venue for further research.

\acknowledgments

We benefitted from discussions with J.~Cardy, G.~Grinstein, and
K.~Oerding. 
U.C.T. acknowledges support from the Engineering and Physical Sciences
Research Council (EPSRC) through Grant GR/J78327, and from the
European Commission through a TMR Grant, contract 
No.~ERB FMBI-CT96-1189. 
Z.R. would like to acknowledge the hospitality of the members of the
Theoretical Physics Department of Oxford University as well as the
support from the Hungarian Academy of Sciences (Grant OTKA T 019451),
and from the EC Network (Grant ERB CHRX-CT92-0063).

\end{multicols}

\appendix

\section{Explicit one--loop results for the vertex functions}
 \label{verfun}

In this Appendix, we present a list of the results to one--loop order
in perturbation theory for those vertex function which are required
for the renormalization of the nonequilibrium SSS model defined by
Eqs.~(\ref{soplan})--(\ref{samnoi}), or, equivalently, by the field
theory (\ref{harfun}), (\ref{relfun}), and (\ref{mctfun}). In all the
subsequent expressions, the internal frequency integrations have
already been carried out via the residue theorem. We use the
abbreviation $\int_k \ldots \equiv (2 \pi)^{-d} \int d^dk \ldots$. We
do not explicitly provide the Feynman diagrams themselves, as they are
identical with those of the equilibrium model \cite{sssfth}.

We begin with the two--point vertex functions renormalizing the noise
strengths ${\tilde \lambda}_0$ and ${\tilde D}_0$,
\begin{eqnarray}
	&&\Gamma_{0 \, {\tilde S} {\tilde S}}({\bf q},\omega) =
	- 2 {\tilde \lambda}_0 \Biggl[ 1 + (n - 1) g_0^2 
	{{\tilde D}_0 \over \lambda_0 D_0} \int_k \! 
	{1 \over r_0 + ({\bf q}/2 + {\bf k})^2} \times \nonumber \\
	&&\qquad \times \left( {D_0 ({\bf q}/2 - {\bf k})^2 \over
	(\omega - i \lambda_0 [r_0 + ({\bf q}/2 + {\bf k})^2])^2 
				+ D_0^2 ({\bf q}/2 - {\bf k})^4} + 
	{\lambda_0 [r_0 + ({\bf q}/2 + {\bf k})^2] \over
	[\omega + i D_0 ({\bf q}/2 - {\bf k})^2]^2 + \lambda_0^2
	[r_0 + ({\bf q}/2 + {\bf k})^2]^2} \right) \Biggr] \ ,
 \label{gmtsts} \\
	&&\Gamma_{0 \, {\tilde M} {\tilde M}}({\bf q},\omega) =
	- 2 {\tilde D}_0 q^2 \Biggl[ 1 + 4 g_0^2 
	{{\tilde \lambda}_0^2 \over \lambda_0 {\tilde D}_0} \int_k \!
	({\bf q} \cdot {\bf k})^2 \times \nonumber \\
	&&\qquad \qquad \qquad \qquad \qquad \times \Biggl( 
	{1 \over r_0 + ({\bf q}/2 + {\bf k})^2} {1 \over 
	(\omega - i \lambda_0 [r_0 + ({\bf q}/2 + {\bf k})^2])^2 +
	\lambda_0^2 [r_0 + ({\bf q}/2 - {\bf k})^2]^2} + \nonumber \\
	&&\qquad \qquad \qquad \qquad \qquad \qquad
	+ {1 \over r_0 + ({\bf q}/2 - {\bf k})^2} {1 \over 
	(\omega + i \lambda_0 [r_0 + ({\bf q}/2 - {\bf k})^2])^2 + 
	\lambda_0^2 [r_0 + ({\bf q}/2 + {\bf k})^2]^2} \Biggr) 
							\Biggr]	\ .  	
 \label{gmtmtm}
\end{eqnarray}
Consequently, we have
\begin{eqnarray}
	&&\Gamma_{0 \, {\tilde S} {\tilde S}}({\bf 0},0) =
	- 2 {\tilde \lambda}_0 \left[ 1 + (n - 1) g_0^2 {{\tilde D}_0 
		\over \lambda_0 D_0} \int_k \! {1 \over r_0 + k^2} 
	{1 \over \lambda_0 (r_0 + k^2) + D_0 k^2} \right] \ , 
 \label{g0tsts} \\
	&&{\partial \over \partial q^2} 
	\Gamma_{0 \, {\tilde M} {\tilde M}}({\bf q},0) 
	\Big\vert_{{\bf q} = {\bf 0}} = - 2 {\tilde D}_0 \left[ 1 + 
	{2 \over d} g_0^2 {{\tilde \lambda}_0^2 \over \lambda_0^3 
	{\tilde D}_0} \int_k \! {k^2 \over (r_0 + k^2)^3} \right] \ .
 \label{g0tmtm}
\end{eqnarray}

For the computation of the response functions, one needs the
two--point functions
\begin{eqnarray}
	&&\Gamma_{0 \, {\tilde S} S}({\bf q},\omega) =
	i \omega + \lambda_0 (r_0 + q^2) + {n + 2 \over 6} \,
	{\tilde \lambda}_0 u_0 \int_k \! {1 \over r_0 + k^2} +
	\nonumber \\ &&\qquad \qquad \qquad +
	(n - 1) (r_0 + q^2) g_0^2 {{\tilde \lambda}_0 \over \lambda_0}
		\int_k \! {1 \over r_0 + ({\bf q}/2 + {\bf k})^2} 
	{1 \over i \omega + \lambda_0 [r_0 + ({\bf q}/2 + {\bf k})^2]
			+ D_0 ({\bf q}/2 - {\bf k})^2} + \nonumber \\
	&&\qquad \qquad \qquad + (n - 1) g_0^2 
	{{\tilde D}_0 \over D_0} \left( 1 - {{\tilde \lambda}_0 \over 
		\lambda_0} {D_0 \over {\tilde D}_0} \right) \int_k \! 
	{1 \over i \omega + \lambda_0 [r_0 + ({\bf q}/2 + {\bf k})^2]
			+ D_0 ({\bf q}/2 - {\bf k})^2} \ ,
 \label{gamtss} \\	
	&&\Gamma_{0 \, {\tilde M} M}({\bf q},\omega) = 
	i \omega + D_0 q^2 - 
	4 g_0^2 {{\tilde \lambda}_0 \over \lambda_0} \int_k \! 
	{({\bf q} \cdot {\bf k}) \over r_0 + ({\bf q}/2 + {\bf k})^2} 
	{1 \over i \omega + 2 \lambda_0 (r_0 + q^2/4 + k^2)} \ ;
 \label{gamtmm}
\end{eqnarray}
specifically,
\begin{eqnarray}
	&&\Gamma_{0 \, {\tilde S} S}({\bf 0},0) = 
	\lambda_0 \Biggl[ r_0 \left( 1 + (n - 1) g_0^2 
		{{\tilde \lambda}_0 \over \lambda_0^2} \int_k \! 
	{1 \over r_0 + k^2} {1 \over \lambda_0 (r_0 + k^2) + D_0 k^2} 
		\right) + \nonumber \\ &&\qquad \qquad \qquad \qquad
	+ {n + 2 \over 6} \, {{\tilde \lambda}_0 \over \lambda_0} u_0 
	\int_k \! {1 \over r_0 + k^2} + (n - 1) g_0^2 {{\tilde D}_0 
	\over \lambda_0 D_0} \left( 1 - {{\tilde \lambda}_0 \over 
	\lambda_0} {D_0 \over {\tilde D}_0} \right) \int_k \! 
	{1 \over \lambda_0 (r_0 + k^2) + D_0 k^2} \Biggr] \ , 
 \label{gm0tss} \\
	&&{\partial \over \partial q^2} 
	\Gamma_{0 \, {\tilde M} M}({\bf q},0) 
	\Big\vert_{{\bf q} = {\bf 0}} = D_0 \left[ 1 + {2 \over d} 
		g_0^2 {{\tilde \lambda}_0 \over \lambda_0^2 D_0} 
		\int_k \! {k^2 \over (r_0 + k^2)^3} \right] \ .
 \label{gm0tmm}
\end{eqnarray}
Furthermore, the following vertex functions containing composite
operators are required [note Eq.~(\ref{compop})],
\begin{eqnarray}
	&&\Gamma_{0 \, {\tilde S} [{\tilde M} S]}({\bf q},\omega) =
	- (n - 1) g_0 {{\tilde \lambda}_0 \over \lambda_0} \int_k \!
	{1 \over r_0 + ({\bf q}/2 + {\bf k})^2}
	{1 \over -i \omega + \lambda_0 [r_0 + ({\bf q}/2 + {\bf k})^2]
				+ D_0 ({\bf q}/2 - {\bf k})^2} \ ,
 \label{gtstms} \\
	&&\Gamma_{0 \, {\tilde M} [{\tilde S} S]}({\bf q},\omega) =
	2 g_0 {{\tilde \lambda}_0 \over \lambda_0} \int_k \! 
	{({\bf q} \cdot {\bf k}) \over r_0 + ({\bf q}/2 + {\bf k})^2}
	{1 \over - i \omega + 2 \lambda_0 (r_0 + q^2/4 + k^2)} \ .
 \label{gtmtss}
\end{eqnarray}

Calculating the three-- and four--point functions is already a rather
tedious task even to one--loop order, and we merely quote the final
results needed for renormalization purposes:
\begin{eqnarray}
	&&\Gamma_{0 \, {\tilde S} S M}({\bf 0},0;{\bf 0},0;{\bf 0},0)
	= - g_0 \Biggl[ 1 - (n - 1) g_0^2 {{\tilde D}_0 \over D_0}
	\int_k \! {1 \over [\lambda_0 (r_0 + k^2) + D_0 k^2]^2} +
	\nonumber \\ &&\qquad \qquad \qquad \qquad \qquad \qquad
			\qquad \qquad \qquad \qquad 
	+ (n - 1) g_0^2 {{\tilde \lambda}_0 \over \lambda_0} 
	\int_k \! {k^2 \over r_0 + k^2} 
	{1 \over [\lambda_0 (r_0 + k^2) + D_0 k^2]^2} \Biggr] \ ,
 \label{g0tssm} \\
	&&{\partial \over \partial ({\bf q} \cdot {\bf p})} 
	\Gamma_{0 \, {\tilde M} S S}
	(-{\bf q},0;{\bf q}/2 - {\bf p},0;{\bf q}/2 + {\bf p},0)
	\Big\vert_{{\bf q} = {\bf p} = {\bf 0}} = 2 g_0 \Biggl[ 1 - 
	{2 \over d} (n - 1) g_0^2 {{\tilde D}_0 \over \lambda_0} 
	\int_k \! {k^2 \over r_0 + k^2} 
	{1 \over [\lambda_0 (r_0 + k^2) + D_0 k^2]^2} + \nonumber \\
	&&\qquad \qquad \qquad \qquad \qquad \qquad \qquad \qquad
	\qquad \qquad + {2 \over d} (n - 1) g_0^2 
	{{\tilde \lambda}_0 D_0 \over \lambda_0^2} 
	\int_k \! {k^4 \over (r_0 + k^2)^2} 
	{1 \over [\lambda_0 (r_0 + k^2) + D_0 k^2]^2} \Biggr] \ ;	
 \label{g0tmss}
\end{eqnarray}
and finally
\begin{eqnarray}
	&&\Gamma_{0 \, {\tilde S} S S S}({\bf 0},0;{\bf 0},0;
	{\bf 0},0;{\bf 0},0) = \lambda_0 u_0 \Biggl[ 1 - 
	{n + 8 \over 6} \, {{\tilde \lambda}_0 \over \lambda_0} u_0 
	\int_k \! {1 \over (r_0 + k^2)^2} + \nonumber \\
	&&\qquad \qquad \qquad \qquad \qquad + (n - 1) g_0^2 
	\left( 1 - {3 g_0^2 \over \lambda_0 D_0 u_0} \right)
	{{\tilde D}_0 \over \lambda_0 D_0} \int_k \! {1 \over r_0 + 
	k^2} {1 \over \lambda_0 (r_0 + k^2) + D_0 k^2} - \nonumber \\
	&&\qquad \qquad \qquad \qquad \qquad - (n - 1) g_0^2 
	\left( 1 + {3 g_0^2 \over \lambda_0 D_0 u_0} \right) 
		{{\tilde D}_0 \over D_0} \int_k \! 
	{1 \over [\lambda_0 (r_0 + k^2) + D_0 k^2]^2} + \nonumber \\
	&&\qquad \qquad \qquad \qquad \qquad + (n - 1) g_0^2 
	\left( 1 + {3 g_0^2 \over \lambda_0 D_0 u_0} \right) 
	{{\tilde \lambda}_0 \over \lambda_0} \int_k \! {k^2 \over 
	r_0 + k^2} {1 \over [\lambda_0 (r_0 + k^2) + D_0 k^2]^2} +
	\nonumber \\ &&\qquad \qquad \qquad \qquad \qquad + (n - 1) 
	{3 g_0^4 \over \lambda_0 D_0 u_0} {{\tilde \lambda}_0 \over 
	\lambda_0^2} \int_k \! {k^2 \over (r_0 + k^2)^2} 
	{1 \over \lambda_0 (r_0 + k^2) + D_0 k^2} - \nonumber \\
	&&\qquad \qquad \qquad \qquad \qquad
	- (n - 1) {3 g_0^4 \over \lambda_0 D_0 u_0} 
	{{\tilde D}_0 \over \lambda_0} \int_k \! {k^2 \over r_0 + k^2}
	{1 \over [\lambda_0 (r_0 + k^2) + D_0 k^2]^2} + \nonumber \\
	&&\qquad \qquad \qquad \qquad \qquad + (n - 1) 
	{3 g_0^4 \over \lambda_0 D_0 u_0} {{\tilde \lambda}_0 D_0 
	\over \lambda_0^2} \int_k \! {k^4 \over (r_0 + k^2)^2} 
	{1 \over [\lambda_0 (r_0 + k^2) + D_0 k^2]^2} \Biggr] \ .
 \label{g0tsss}
\end{eqnarray}

\begin{multicols}{2}

\end{multicols}

\vskip 1 truecm

{\bf FIGURE CAPTION:}

\begin{figure}
\caption{One--loop flow diagram for the nonequilibrium SSS model in
	the space of dynamical couplings $w / (1 + w)$, 
	${\bar f} / (1 + {\bar f})$, and ${\tilde f} / 2$ (displayed 
	for the case $n = 2$, $\epsilon = 1$).
	Asymptotically, the equilibrium strong--scaling fixed point 
	$w_{\rm eq}^*= 2 n - 3$, ${\tilde f}_{\rm eq}^* = \epsilon$, 
	${\bar f}_{\rm eq}^* = \epsilon / (2 n - 3)$ is stable; in the
	figure, it is located at the point $(1/2,1/2,1/2)$ 
	(full circle).}
 \label{flodia}
\end{figure}

\newpage

\begin{figure}
\epsfxsize = 5.5 truein
\epsffile{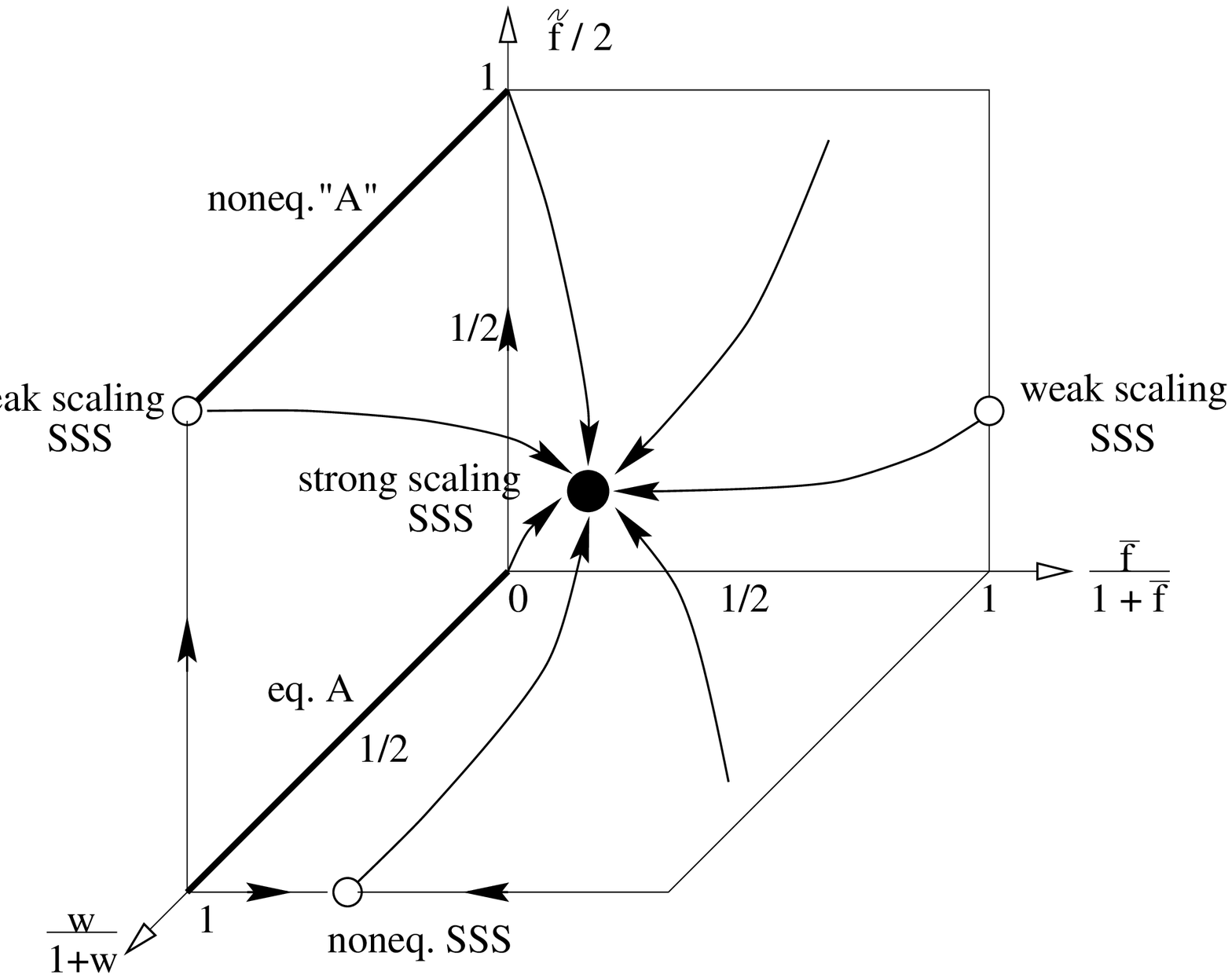}
\end{figure}

\end{document}